\documentclass[aps,prl,twocolumn,amsmath,amssymb,nofootinbib,superscriptaddress]{revtex4-1}

\usepackage{times}
\usepackage[pdftex]{graphicx}
\usepackage{dcolumn}
\usepackage{bm}
\usepackage{amsmath}
\usepackage{indentfirst}
\usepackage{float}
\usepackage[colorlinks]{hyperref}
\usepackage[dvipsnames]{xcolor}
\usepackage{xcolor}

\usepackage{subfigure}
\usepackage{algorithm}
\usepackage{algorithmicx}
\usepackage{algpseudocode}
\usepackage{amsmath}
\usepackage{verbatim}
\usepackage{makecell}

\begin{document}

\title{Quantum Computational Advantage via 60-Qubit 24-Cycle Random Circuit Sampling}
%Enhancing Quantum Computational Advantage via 60-Qubit Random Circuit Sampling}
%Characterizing a 66-Qubit Quantum Processor via Random Quantum Circuit Benchmarking
%60-Qubit
%Enhanced Quantum Computational Advantage using a Superconducting Quantum Processor
%\author{XXX}
%\affiliation{Hefei National Laboratory for Physical Sciences at the Microscale and Department of Modern Physics, University of Science and Technology of China, Hefei 230026, China}
%\affiliation{Shanghai Branch, CAS Center for Excellence in Quantum Information and Quantum Physics, University of Science and Technology of China, Shanghai 201315, China}
%\affiliation{Shanghai Research Center for Quantum Sciences, Shanghai 201315, China}

\author{Qingling Zhu}
\affiliation{Hefei National Laboratory for Physical Sciences at the Microscale and Department of Modern Physics, University of Science and Technology of China, Hefei 230026, China}
\affiliation{Shanghai Branch, CAS Center for Excellence in Quantum Information and Quantum Physics, University of Science and Technology of China, Shanghai 201315, China}
\affiliation{Shanghai Research Center for Quantum Sciences, Shanghai 201315, China}
\author{Sirui Cao}
\author{Fusheng Chen}
\author{Ming-Cheng Chen}
\affiliation{Hefei National Laboratory for Physical Sciences at the Microscale and Department of Modern Physics, University of Science and Technology of China, Hefei 230026, China}
\affiliation{Shanghai Branch, CAS Center for Excellence in Quantum Information and Quantum Physics, University of Science and Technology of China, Shanghai 201315, China}
\affiliation{Shanghai Research Center for Quantum Sciences, Shanghai 201315, China}
\author{Xiawei Chen}
\affiliation{Shanghai Branch, CAS Center for Excellence in Quantum Information and Quantum Physics, University of Science and Technology of China, Shanghai 201315, China}
\author{Tung-Hsun Chung}
\author{Hui Deng}
\affiliation{Hefei National Laboratory for Physical Sciences at the Microscale and Department of Modern Physics, University of Science and Technology of China, Hefei 230026, China}
\affiliation{Shanghai Branch, CAS Center for Excellence in Quantum Information and Quantum Physics, University of Science and Technology of China, Shanghai 201315, China}
\affiliation{Shanghai Research Center for Quantum Sciences, Shanghai 201315, China}
\author{Yajie Du}
\affiliation{Shanghai Branch, CAS Center for Excellence in Quantum Information and Quantum Physics, University of Science and Technology of China, Shanghai 201315, China}
\author{Daojin Fan}
\author{Ming Gong}
\author{Cheng Guo}
\author{Chu Guo}
\author{Shaojun Guo}
\author{Lianchen Han}
\affiliation{Hefei National Laboratory for Physical Sciences at the Microscale and Department of Modern Physics, University of Science and Technology of China, Hefei 230026, China}
\affiliation{Shanghai Branch, CAS Center for Excellence in Quantum Information and Quantum Physics, University of Science and Technology of China, Shanghai 201315, China}
\affiliation{Shanghai Research Center for Quantum Sciences, Shanghai 201315, China}
\author{Linyin Hong}
\affiliation{QuantumCTek Co., Ltd., Hefei 230026, China}
\author{He-Liang Huang}
\affiliation{Hefei National Laboratory for Physical Sciences at the Microscale and Department of Modern Physics, University of Science and Technology of China, Hefei 230026, China}
\affiliation{Shanghai Branch, CAS Center for Excellence in Quantum Information and Quantum Physics, University of Science and Technology of China, Shanghai 201315, China}
\affiliation{Shanghai Research Center for Quantum Sciences, Shanghai 201315, China}
\affiliation{Henan Key Laboratory of Quantum Information and Cryptography, Zhengzhou 450000, China}
\author{Yong-Heng Huo}
\affiliation{Hefei National Laboratory for Physical Sciences at the Microscale and Department of Modern Physics, University of Science and Technology of China, Hefei 230026, China}
\affiliation{Shanghai Branch, CAS Center for Excellence in Quantum Information and Quantum Physics, University of Science and Technology of China, Shanghai 201315, China}
\affiliation{Shanghai Research Center for Quantum Sciences, Shanghai 201315, China}
\author{Liping Li}
\affiliation{Shanghai Branch, CAS Center for Excellence in Quantum Information and Quantum Physics, University of Science and Technology of China, Shanghai 201315, China}
\author{Na Li}
\author{Shaowei Li}
\author{Yuan Li}
\author{Futian Liang}
\affiliation{Hefei National Laboratory for Physical Sciences at the Microscale and Department of Modern Physics, University of Science and Technology of China, Hefei 230026, China}
\affiliation{Shanghai Branch, CAS Center for Excellence in Quantum Information and Quantum Physics, University of Science and Technology of China, Shanghai 201315, China}
\affiliation{Shanghai Research Center for Quantum Sciences, Shanghai 201315, China}
\author{Chun Lin}
\affiliation{Shanghai Institute of Technical Physics, Chinese Academy of Sciences, Shanghai 200083, China}
\author{Jin Lin}
\author{Haoran Qian}
\affiliation{Hefei National Laboratory for Physical Sciences at the Microscale and Department of Modern Physics, University of Science and Technology of China, Hefei 230026, China}
\affiliation{Shanghai Branch, CAS Center for Excellence in Quantum Information and Quantum Physics, University of Science and Technology of China, Shanghai 201315, China}
\affiliation{Shanghai Research Center for Quantum Sciences, Shanghai 201315, China}
\author{Dan Qiao}
\affiliation{Shanghai Branch, CAS Center for Excellence in Quantum Information and Quantum Physics, University of Science and Technology of China, Shanghai 201315, China}
\author{Hao Rong}
\author{Hong Su}
\author{Lihua Sun}
\affiliation{Hefei National Laboratory for Physical Sciences at the Microscale and Department of Modern Physics, University of Science and Technology of China, Hefei 230026, China}
\affiliation{Shanghai Branch, CAS Center for Excellence in Quantum Information and Quantum Physics, University of Science and Technology of China, Shanghai 201315, China}
\affiliation{Shanghai Research Center for Quantum Sciences, Shanghai 201315, China}
\author{Liangyuan Wang}
\affiliation{Shanghai Branch, CAS Center for Excellence in Quantum Information and Quantum Physics, University of Science and Technology of China, Shanghai 201315, China}
\author{Shiyu Wang}
\author{Dachao Wu}
\author{Yulin Wu}
\author{Yu Xu}
\affiliation{Hefei National Laboratory for Physical Sciences at the Microscale and Department of Modern Physics, University of Science and Technology of China, Hefei 230026, China}
\affiliation{Shanghai Branch, CAS Center for Excellence in Quantum Information and Quantum Physics, University of Science and Technology of China, Shanghai 201315, China}
\affiliation{Shanghai Research Center for Quantum Sciences, Shanghai 201315, China}
\author{Kai Yan}
\affiliation{Shanghai Branch, CAS Center for Excellence in Quantum Information and Quantum Physics, University of Science and Technology of China, Shanghai 201315, China}
\author{Weifeng Yang}
\affiliation{QuantumCTek Co., Ltd., Hefei 230026, China}
\author{Yang Yang}
\affiliation{Shanghai Branch, CAS Center for Excellence in Quantum Information and Quantum Physics, University of Science and Technology of China, Shanghai 201315, China}
\author{Yangsen Ye}
\affiliation{Hefei National Laboratory for Physical Sciences at the Microscale and Department of Modern Physics, University of Science and Technology of China, Hefei 230026, China}
\affiliation{Shanghai Branch, CAS Center for Excellence in Quantum Information and Quantum Physics, University of Science and Technology of China, Shanghai 201315, China}
\affiliation{Shanghai Research Center for Quantum Sciences, Shanghai 201315, China}
\author{Jianghan Yin}
\affiliation{Shanghai Branch, CAS Center for Excellence in Quantum Information and Quantum Physics, University of Science and Technology of China, Shanghai 201315, China}
\author{Chong Ying}
\author{Jiale Yu}
\author{Chen Zha}
\author{Cha Zhang}
\affiliation{Hefei National Laboratory for Physical Sciences at the Microscale and Department of Modern Physics, University of Science and Technology of China, Hefei 230026, China}
\affiliation{Shanghai Branch, CAS Center for Excellence in Quantum Information and Quantum Physics, University of Science and Technology of China, Shanghai 201315, China}
\affiliation{Shanghai Research Center for Quantum Sciences, Shanghai 201315, China}
\author{Haibin Zhang}
\affiliation{Shanghai Branch, CAS Center for Excellence in Quantum Information and Quantum Physics, University of Science and Technology of China, Shanghai 201315, China}
\author{Kaili Zhang}
\author{Yiming Zhang}
\affiliation{Hefei National Laboratory for Physical Sciences at the Microscale and Department of Modern Physics, University of Science and Technology of China, Hefei 230026, China}
\affiliation{Shanghai Branch, CAS Center for Excellence in Quantum Information and Quantum Physics, University of Science and Technology of China, Shanghai 201315, China}
\affiliation{Shanghai Research Center for Quantum Sciences, Shanghai 201315, China}
\author{Han Zhao}
\affiliation{Shanghai Branch, CAS Center for Excellence in Quantum Information and Quantum Physics, University of Science and Technology of China, Shanghai 201315, China}
\author{Youwei Zhao}
\affiliation{Hefei National Laboratory for Physical Sciences at the Microscale and Department of Modern Physics, University of Science and Technology of China, Hefei 230026, China}
\affiliation{Shanghai Branch, CAS Center for Excellence in Quantum Information and Quantum Physics, University of Science and Technology of China, Shanghai 201315, China}
\affiliation{Shanghai Research Center for Quantum Sciences, Shanghai 201315, China}
\author{Liang Zhou}
\affiliation{QuantumCTek Co., Ltd., Hefei 230026, China}
\author{Chao-Yang Lu}
\author{Cheng-Zhi Peng}
\author{Xiaobo Zhu}
\author{Jian-Wei Pan}

\affiliation{Hefei National Laboratory for Physical Sciences at the Microscale and Department of Modern Physics, University of Science and Technology of China, Hefei 230026, China}
\affiliation{Shanghai Branch, CAS Center for Excellence in Quantum Information and Quantum Physics, University of Science and Technology of China, Shanghai 201315, China}
\affiliation{Shanghai Research Center for Quantum Sciences, Shanghai 201315, China}

%\date{\today}


\begin{abstract}
To ensure a long-term quantum computational advantage, the quantum hardware should be upgraded to withstand the competition of continuously improved classical algorithms and hardwares. Here, we demonstrate a superconducting quantum computing systems \textit{Zuchongzhi} 2.1, which has 66 qubits in a two-dimensional array in a tunable coupler architecture.  The readout fidelity of  \textit{Zuchongzhi} 2.1 is considerably improved to an average of 97.74\%. The more powerful quantum processor enables us to achieve larger-scale random quantum circuit sampling, with a system scale of up to 60 qubits and 24 cycles. The achieved sampling task is about 6 orders of magnitude more difficult than that of Sycamore [Nature \textbf{574}, 505 (2019)] in the classic simulation, and 3 orders of magnitude more difficult than the sampling task on \textit{Zuchongzhi} 2.0 [arXiv:2106.14734 (2021)]. The time consumption of classically simulating random circuit sampling experiment using state-of-the-art classical algorithm and supercomputer is extended to tens of thousands of years (about $4.8\times 10^4$ years), while \textit{Zuchongzhi} 2.1 only takes about 4.2 hours, thereby significantly enhancing the quantum computational advantage.
\end{abstract}

\maketitle
%\newpage

\section{Introduction}

Quantum computers are emerging with the promise of solving certain computational tasks exponentially faster than the current classical computers~\cite{nielsen2002quantum, shor1999polynomial}. Limited by the development of quantum hardwares, quantum computing has been at the stage of small-scale demonstrations for several decades in the past~\cite{huang2020superconducting}. Recently, thanks to the significant progress in superconducting and photonic platforms, the long-anticipated milestone of quantum computational advantage or supremacy~\cite{aaronson2011computational,boixo2018characterizing,bouland2019complexity,aaronson2016complexity} has been achieved by using the quantum processors Sycamore~\cite{arute2019quantum}, \textit{Jiuzhang}~\cite{zhong2020quantum,zhong2021phase}, and \textit{Zuchongzhi}~\cite{wu2021strong} successively. These quantum processors with dozens of qubits reveal remarkable potential for quantum computing to offer new capabilities for near-term applications in the noisy intermediate scale quantum (NISQ) area, such as quantum simulation~\cite{google2020hartree,mcardle2020quantum,aspuru2005simulated, bernien2017probing,zhang2017observation, zhu2021observation,chen2021observation,zha2020ergodic,yang2020quantum,cian2021many,xu2018emulating,gong2021quantum,huang2021emulating,liu2019demonstration}, quantum machine learning~\cite{huang2020experimental,havlivcek2019supervised,harrigan2021quantum,huang2018demonstration, saggio2021experimental,cong2019quantum,liu2021hybrid}, and cloud quantum computing~\cite{barz2012demonstration,huang2017experimental,huang2017universal,dumitrescu2018cloud,huang2017homomorphic,huang2018entanglement}.

Nonetheless, quantum computational advantage will be a long-term competition between classical simulation and quantum devices, rather than being a one-off experimental demonstration. During the past two years, the classical simulation algorithms keep evolving~\cite{pednault2019leveraging,HuangChen2020,pan2021simulating,MarkovShi2008,GuoWu2019,VillalongaMandra2019,villalonga2020establishing,guo2021verifying}. In particular, it is estimated in Ref.~\cite{HuangChen2020} that simulating Sycamore's 53-qubit and 20-cycle circuit sampling only takes about 19 days on Summit, and the exact amplitudes for 2 million correlated bitstrings were computed in 5 days by using only 60 GPUs in a recent work~\cite{pan2021simulating}, in comparison with 10,000 years as estimated in the original paper~\cite{arute2019quantum}. Additionally, since calculating and storing the evolution of the entire state-space of the 53 qubits in random circuit sampling is still possible~\cite{pednault2019leveraging}, there is a sample-size-dependent loophole in the Google's Sycamore experiment. The development of classical algorithms and hardwares is almost about to overturn the quantum computational advantage that Sycamore has achieved. Thus, the quantum hardware needs to be continuously upgraded to maintain the quantum computational advantage. Such a long-term competition will also provide an important diagnose to benchmark the quantum computing system which is useful for eventually realizing fault-tolerant quantum computing.

In this work, a new 66-qubit two-dimensional superconducting quantum processor, called \textit{Zuchongzhi} 2.1, is designed and fabricated. Compared to \textit{Zuchongzhi} 2.0~\cite{wu2021strong}, the readout performance of \textit{Zuchongzhi} 2.1 is greatly improved to an average fidelity of 97.74\%, while the single-qubit gate and the two-qubit gate have an average fidelity of 99.84\% and 99.40\%, respectively. Based on this state-of-the-art quantum processor, the random quantum circuit sampling with a system size up to 60 qubits and 24 cycles is realized experimentally. As estimated, the classical computational cost of simulating our sample task is about 6 orders of magnitude and 5000 times higher than that of the hardest tasks on the Sycamore ~\cite{arute2019quantum} and \textit{Zuchongzhi} 2.0 processors, respectively. Our experiment extendeds the classical simulation time of random circuit sampling experiment to tens of thousands of years, even using the state-of-the-art classical supercomputers.

\section{\textit{Zuchongzhi} 2.1 —— an Upgraded High-Performance Quantum Processor}

\begin{figure}[!htbp]
\begin{center}
\includegraphics[width=0.95\linewidth]{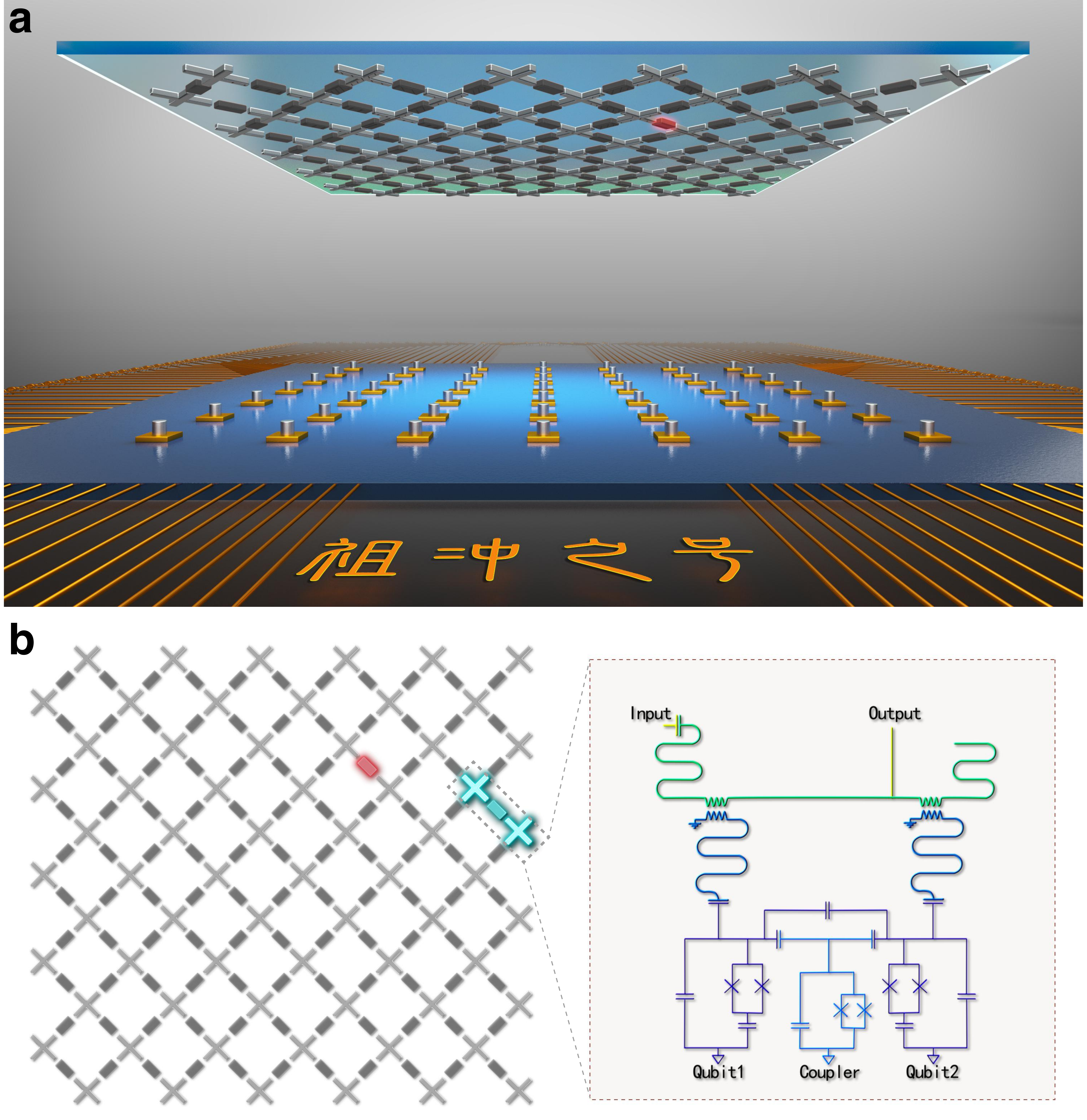}
\end{center}
\caption{\textbf{Device schematic of the \textit{Zuchongzhi} 2.1 quantum processor.} (a) The  \textit{Zuchongzhi} 2.1 quantum processor is composed of two sapphire chips combined by flip-chip technique, one of which has 66 qubits and 110 couplers, and the other carries all the control lines. (b) Simplified circuit schematic of the qubit, coupler and readout. The coupler marked in red in the left panel is not working properly, which can only provide fixed coupling with a coupling strength of about 7.5 MHz.}
\label{fig1}
\end{figure}

\begin{figure*}[!htbp]
\begin{center}
\includegraphics[width=0.95\linewidth]{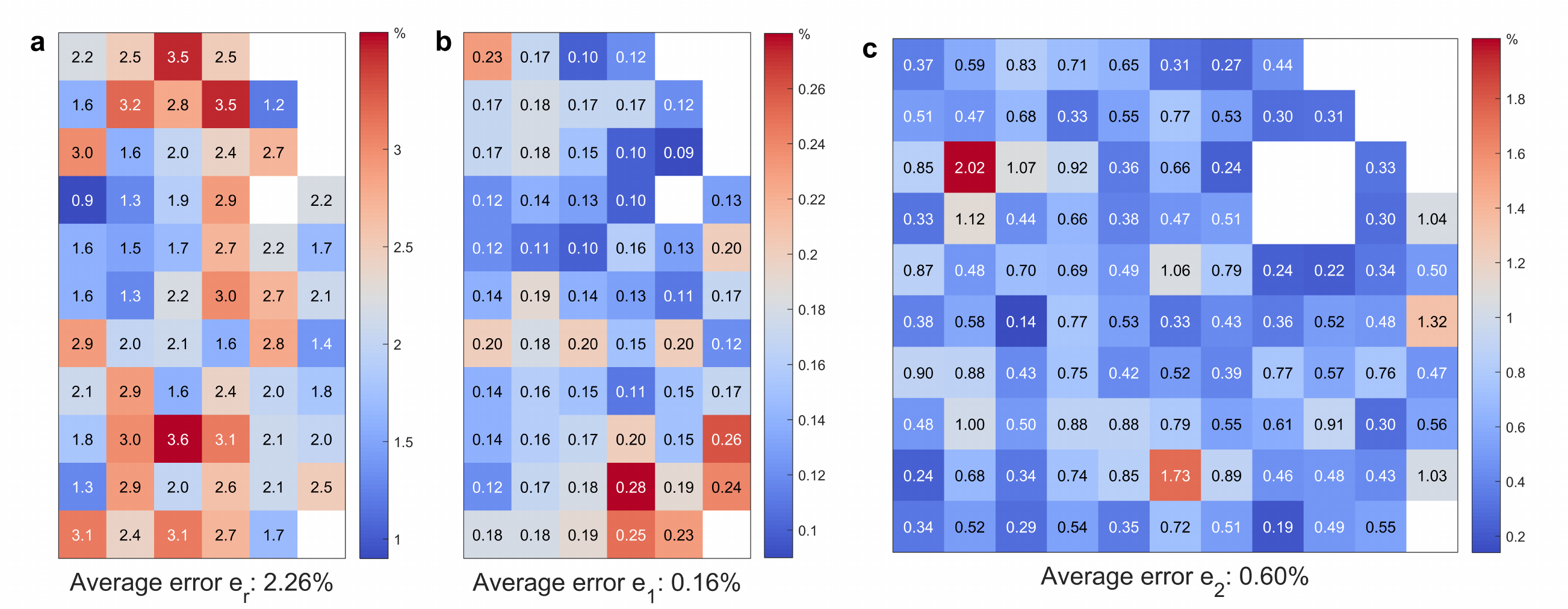}
\end{center}
\caption{\textbf{Readout, single-qubit gate and two-qubit gate performance of the selected 60 qubits.} (a) and (b) are the readout error $e_{r}$ and single-qubit gate pauli error $e_{1}$ for these 60 qubits, respectively. (c) The two qubit gate pauli error $e_{2}$ of the 99 two-qubit gates used. The errors are measured when operating all qubits simultaneously. The average errors of readout and quantum gates are also given in each subgraph.}
\label{fig2}
\end{figure*}

\textit{Zuchongzhi} 2.1 is an upgraded version of the same generation of \textit{Zuchongzhi} 2.0~\cite{wu2021strong} (\textit{Zuchongzhi} 1.0 is used in Ref.~\cite{gong2021quantum}). \textit{Zuchongzhi} 2.1 is designed and fabricated as a $11\times6$ two-dimensional rectangular superconducting qubit array composed of 66 Transmon qubits~\cite{koch:042319} (see Fig.~\ref{fig1}). Except for the qubits at the boundary, each qubit has four tunable couplers to couple to its nearest neighboring qubits~\cite{yysCouplerCZ}, where the coupler is also a Transmon qubit, whose frequency several GHz higher than that of the data qubits~\cite{yan2018tunable}. All the qubits and couplers are fabricated on a sapphire chip, and all the control lines are fabricated on another sapphire chip. These two separate sapphire chips are then stacked together with the indium bump flip-chip technique. Each qubit has a microwave drive line and a fast flux-bias line to realize full control, including state excitation and qubit frequency modulation. Each coupler is controlled by a magnetic flux bias line, to realize fast tuning of the coupling strength $g$ continually from $\sim+5~\text{MHz}$ to $\sim-50~\text{MHz}$. Except for a unfunctional coupler (see Fig.~\ref{fig1}(b)), the remaining 66 qubits and 109 couplers on the quantum processor function properly, among which 60 qubits are selected for random quantum circuit sampling experiment.

Compared with previous work~\cite{wu2021strong},  \textit{Zuchongzhi} 2.1 mainly upgrades the performance of readout. Specifically, the coupler frequency at turn-off point is significantly higher than the readout frequency, so that we don’t need to reset the qubit frequency when performing measurement. After calibrating and optimizing the readout, the average single-qubit state readout error of the selected 60 qubits can reach $e_r=2.26\%$ (Fig.~\ref{fig2}(a)). Among them, the highest readout fidelity reach 99.1\%. The huge improvement in readout performance plays a key role in our realization of larger and deeper random quantum circuit sampling.

The procedures of calibration and optimization of single-qubit and two-qubit gate are similar to \textit{Zuchongzhi} 2.0~\cite{wu2021strong}. In order to achieve parallel execution of gates, all the couplers are turned off when single-qubit gates are applied to isolate each qubit, and the required couplers are turned on only when two-qubit gates are implemented. The two-qubit gate implemented in our experiment is iSWAP-like gate~\cite{arute2019quantum}, for the purpose of random circuit sampling. The matrix form of iSWAP-like gate is:
	\begin{equation}
		\begin{bmatrix}
			1 	& 0 											& 0												& 0 \\
			0 	& e^{i(\Delta_{+}+\Delta_{-})}\rm{cos}\theta 		& -ie^{i(\Delta_{+}-\Delta_{-,\text{off}})}\rm{sin}\theta	& 0 \\
			0 	& -ie^{i(\Delta_{+}+\Delta_{-,\text{off}})}\rm{sin}\theta	& e^{i(\Delta_{+}-\Delta_{-})}\rm{cos}\theta			& 0 \\
			0 	& 0 											& 0												& e^{i(2\Delta_{+}-\phi)}
		\end{bmatrix},
	\end{equation}
where the five parameters $\{\theta, \phi, \Delta_{+}, \Delta_{-}, \Delta_{-,\text{off}}\}$ are obtained by two-qubit gate cross-entropy benchmarking (XEB) ~\cite{boixo2018characterizing, Neill195} fitting.
The use of a stronger coupling strength ($\sim 14~\text{MHz}$) than that ($\sim 10~\text{MHz}$) in 2.0, the time duration of iSWAP-like gate is shorten from $\sim 32~\rm{\text{ns}}$ to $\sim 24~\rm{\text{ns}}$. The performance of quantum gates is characterized by parallel XEB, yielding the average pauli errors of 0.16\% (Fig.~\ref{fig2}(b)) and 0.60\% (Fig.~\ref{fig2}(c)) for single-qubit gates and two-qubit gates, respectively, in the case of all gates are applied simultaneously.

\begin{figure*}[!htbp]
\begin{center}
\includegraphics[width=\linewidth]{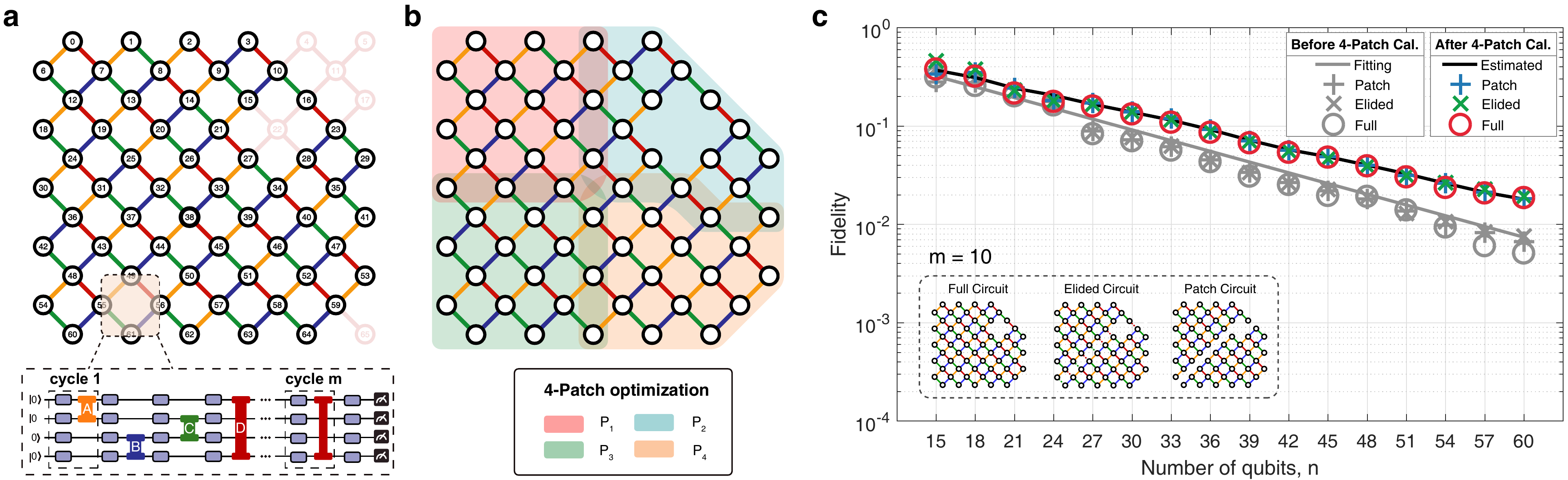}
\end{center}
\caption{\textbf{60-qubit random quantum circuit, 4-patch calibration, and results of cirucit with 15-60 qubits and 10-cycle.} (a) In the processor structure topology diagram, the circles represent qubits, and the discarded qubits are marked with a shaded colour. The single-qubit gates in the random quantum circuit are chosen randomly from the set of $\{ \sqrt X ,\sqrt Y ,\sqrt W \}$. The orange, blue, green, and red lines represent the two-qubit gates of the four patterns A, B, C, and D respectively. (b) We divide the circuit into 4 patches as shown, and optimize the parameters of iSWAP gates in each patch. (c) The XEB fidelities of cirucit with 15-60 qubits and 10-cycle before and after 4-patch calibration. Each data point, including the results from full circuit, patch circuit, and elided circuit, is an average over nine quantum circuit instances. The predicted fidelity result is shown as a black line labeled by “Estimated”, which is calculated from a simple multiplication of individual operation. The gray line labeled by “Fitting” is the exponential fit of the fidelities before 4-patch calibration. Whether it’s before or after 4-patch calibration, the fidelities of the patch and elided circuits are in good agreement with those of the full circuit. And after the 4-patch calibration, the fidelity of each circuit can be significantly improved, and is highly consistent with the predicted fidelity.}
\label{fig3}
\end{figure*}

\begin{figure*}[!htbp]
\begin{center}
\includegraphics[width=0.98\linewidth]{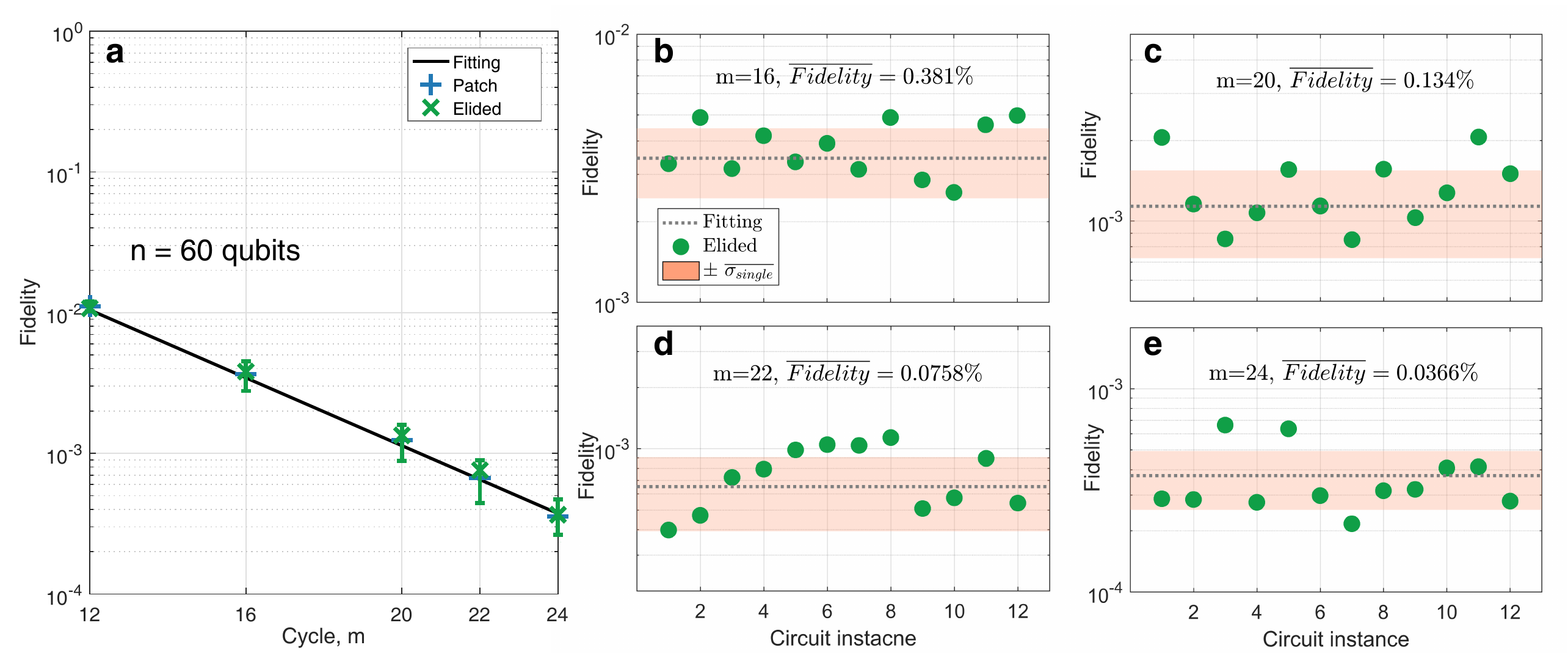}
\end{center}
 \caption{\textbf{Experimental results of  random quantum circuits in the quantum computational advantage regime.} (a) Results of random quantum circuits with 60 qubits and 12-24 cycles. Each data point denotes the average fidelity of the patch or elided circuit over 12 distinct quantum circuit instances, as an estimate of the fidelity of the full circuit. The error bar denotes three standard deviations. For each 60-qubit 24-cycle circuit instance, about 70 million bitstrings are sampled in 4.2 hours (1 million samples per 210 seconds). (b), (c), (d), and (e) show the fidelity of each individual 16-cycle, 20-cycle, 22-cycle, and 22-cycle elided circuit instance, respectively. The band corresponding to $\pm \overline{\sigma_{\text{single}}}$, which is the mean statistical error bar of the 12 instances.}
\label{fig4}
\end{figure*}

\section{60-Qubit Random Circuit Sampling}

After the basic calibration, we implement random quantum circuit sampling to characterize the overall performance of the quantum processor, and measure the degree of quantum computational advantage. The gate sequence of our random quantum circuit is shown in Fig.~\ref{fig3}(a). Each cycle consists of a layer of single-qubit gates randomly selected from set $\{ \sqrt X ,\sqrt Y ,\sqrt W \}$ and applied to all qubits, followed by a layer of two-qubit  iSWAP-like gates. The two-qubit gates in each two-qubit gate layer are applied according to a specified pattern, labeled by A, B, C, and D, and the layer is implemented in the sequence of ABCDCDAB in each cycle.

In general, the verification of large-scale random quantum circuit is done through two variant circuits, named patch circuit and elided circuit, which are designed by removing a slice of two-qubit gates and only a fraction of the gates between the patches, respectively. As the first step of the experiment, we first evaluate whether these two verification methods are applicable on this quantum processor. We implement these two variant circuits and full version of the circuits ranging from 15 qubits to 60 qubits with 10 cycles, and calculate the linear XEB fidelities $F_{\text{XEB}}$ of the respective output bitstrings. Experimental results (see Fig.~\ref{fig3}(c)) show that the average ratio of patch circuit and elided circuit fidelities to full circuit fidelity are 1.09 and 1.13, indicating the effectiveness of these two verification circuits.

However, the fidelity of all circuits appears lower than the predicted fidelity calculated from a simple multiplication of individual operations. This is due to the additional error caused by our control method. In \textit{Zuchongzhi} 2.1, the coupler and qubit are placed at their original frequencies most of the time, and then detuned to the idle frequencies within a few microseconds before and during the experiment, and finally detuned back to their original frequencies after the readout pulse ends (\textit{Zuchongzhi} 2.0 uses DC pulse to keep the qubit and coupler at the idle frequencies all the time.). The distortion of the square pulse causes qubit frequency and coupling strength continuously change slightly during the operation time. To mitigate this effect, the square pulses on the qubit are calibrated. However, the calibration of square pulses on coupler is not performed, due to the frequency of the couplers is too high. This causes the parameters of iSWAP-like gate obtained by the two-qubit XEB to be slightly inaccurate for shallow circuits.

We develop a new calibration method, called 4-patch calibration, to search the optimal parameters of the iSWAP-like gate, based on the original gate parameters obtained by the two-qubit XEB. We divide the full circuit into $4$ non-overlapping patches, which nevertheless covers all the iSWAP gates.  Then we sample the bitstrings of the 4 patch circuits separately, and use the sampled bitstrings as the training set $b_{\text{train}}$. For each patch, we use a gradient-based machine learning to optimize the parameters of the iSWAP gates it contains. Specifically, we consider the set $\gamma {\rm{ = \{ }}{\vec \gamma _1},...,{\vec \gamma _{ng}}{\rm{\} }}$ as trainable parameters, where $\vec\gamma _i=({\theta}^i, {\phi}^i, {\Delta_{+}}^i, {\Delta_{-}}^i, {\Delta_{-,\text{off}} }^i)$ is the parameters of $i$-th iSWAP gate, and $ng$ is the total number of iSWAP gates in this patch. By minimizing the loss function
\begin{align}\label{eq:loss}
\text{Loss}_{\text{XEB}}({\gamma}, b_{\text{train}}) = 1 - F_{\text{XEB}}({\gamma}, b_{\text{train}}),
\end{align}
using the BFGS optimizer, we are able to find the optimal parameters of each iSWAP for this patch, where $F_{\text{XEB}}({\gamma}, b_{\text{train}})$ is the linear XEB fidelity of chosen patch. All these trainings are implemented on a classical computer, and thus we can compute the gradient of Eq.(\ref{eq:loss}) using the classical auto-differentiation framework. By separately training the parameters of the iSWAP gate in the 4 patches, the optimized parameters of all the iSWAP gates are obtained. After 4-patch calibration, we re-calculate the XEB fidelities of all of the circuits in Fig.~\ref{fig3}(c) using the the optimized two-qubit gate parameters. Results show that all of these fidelities have been greatly improved and match the predicted fidelities quite well (see Fig.~\ref{fig3}(c)), indicating that the optimized parameters have adaptability to different quantum circuits. Furthermore, The behavior of two variant circuits are more consistent with full circuits, where the the average ratio of patch circuit and elided circuit fidelities to full circuit fidelity are improved to 1.05 and 1.07, respectively. Since the cirucit is split into 4 patches for training, our 4-patch calibration method can be efficiently implemented on the classical computer. For larger circuits, our method can easily be modified into $n$-patch calibration to maintain its efficiency.

\begin{table*}[!htbp]
\caption{The runtime of tensor network algorithm for different circuits on Summit.
The classical simulation consumption estimation of the random quantum circuit sampling experiment on the Sycamore, \textit{Zuchongzhi} 2.0, and \textit{Zuchongzhi} 2.1 processors are provided. FPOs is the abbreviation for the number of floating point operations, QPU is the abbreviation for quantum processing unit.}
\setlength{\tabcolsep}{0.8mm}{\begin{tabular}[c]{@{}ccccccccc}
\toprule
Processor & Circuit & Fidelity & \# of bitstrings    &
\makecell[c]{FPOs\\ (a perfect sample)}& \makecell[c]{FPOs\\ (circuit)} & \makecell[c]{Runtime on \\Summit} & \makecell[c]{Runtime on\\QPU} &$\frac{{{\rm{Classical  Runtime}}}}{{{\rm{Qauntum  Runtime}}}}$\\
\colrule
Sycamore~\cite{arute2019quantum} &53-qubit 20-cycle & $0.224\%$ &  $3.0 \times 10^6$ & $1.63\times 10^{18}$ & $1.10\times 10^{22}$  & 15.9 days&600s& $2.29\times10^3$\\
\textit{Zuchongzhi} 2.0~\cite{wu2021strong} & 56-qubit 20-cycle & $0.0662\%$ &  $1.9 \times 10^7$ & $1.65\times 10^{20}$ & $2.08\times 10^{24}$  & 8.2 years&1.2h& $6.02\times10^4$\\
\textit{Zuchongzhi} 2.1 & 60-qubit 22-cycle & $0.0758\%$ &  $1.5 \times 10^7$ & $1.06\times 10^{22}$ & $1.21\times 10^{26}$  & $4.8\times 10^2$ years & 1h&$4.21\times10^6$\\
\textit{Zuchongzhi} 2.1 & 60-qubit 24-cycle & $0.0366\%$ &  $7.0 \times 10^7$ & $4.68\times 10^{23}$ & $1.2\times 10^{28}$  & $4.8\times 10^4$ years &4.2h&$9.93\times10^7$\\
\botrule
\end{tabular}}
\label{tab:TNC}
\end{table*}

We sample the output bitstrings of 60-qubit full, patch, and elided circuits with increasing depth from 12 to 24 cycles , and employ patch and elided circuits to estimate the performance of full circuit. At each system size, a total of 12 randomly generated circuit instances are executed. Figure~\ref{fig4}(a) shows the XEB fidelities of patch and elided circuits. For the largest circuit with 60 qubits and 24 cycles, we collect approximately $7.0 \times {10^7}$ bitstrings for each circuit instance, and the combined linear XEB fidelity of theses 12 elided circuit instances are calculated as $F_{\text{XEB}}=(3.66 \pm 0.345)\times 10^{-4}$. Thus, with high statistical significance ($>10\sigma$), the null hypothesis of uniform sampling ($F_{\text{XEB}}=0$) is rejected.
%In addition, each individual circuit instance fidelity is nearly inside the $\pm \overline{\sigma_{\text{single}}}$ statistical error band as shown in Fig.~\ref{fig4}(b-e), where $\overline{\sigma_{\text{single}}}=\frac{1}{{12}}\sum\limits_{i = 1}^{12} {{\sigma _i}}$ is the average statistical error of 12 single instances, indicating the stability of the system and the unbiasedness of noise.

\section{Computational Cost Estimation}

We employ the state-of-the-art classical algorithm, tensor network algorithm~\cite{MarkovShi2008,GuoWu2019,VillalongaMandra2019,villalonga2020establishing,HuangChen2020,pan2021simulating,guo2021verifying}, to estimate the classical computational cost of our largest circuit with 60 qubits and 24 cycles. The number of floating point operations to generate one perfect sample from the $60$-qubit 24-cycle random circuit are estimated as $1.63\times 10^{18}$ by the python package cotengra~\cite{cotengra} (see Supplemental Material for details). Thus, it would cost a total of $1.10\times 10^{22}$ floating point operations to reproduce the same results of this hardest circuit ($7.0 \times {10^7}$ bitstrings, 0.0366\% fidelity) using classical computer. Table~\ref{tab:TNC} also shows the extrapolated run times for Sycamore's 53-qubit 20-cycle circuit ~\cite{arute2019quantum} and \textit{Zuchongzhi} 2.0's 56-qubit 20-cycle circuit using tensor network algorithm.

As shown in Table~\ref{tab:TNC}, using the tensor network algorithm, the classical computational cost of our sample task with 60-qubit and 24-cycle is about 6 orders of magnitude and 5000 times higher than that of the hardest tasks on the Sycamore and \textit{Zuchongzhi} 2.0 processor~\cite{wu2021strong}, respectively. The random circuit sampling task with 60-qubit and 24-cycles, achieved by \textit{Zuchongzhi} 2.1 in 4.2 hours, would cost Summit 48,000 years to simulate.

%We also used the Schr$\ddot{\text{o}}$dinger-Feynman algorithm (SFA), a state-of-the-art full-amplitude algorithm, to evaluate the classical computational cost. But at this scale, the SFA algorithm is far less efficient than the tensor algorithm. The results of applying Schr$\ddot{\text{o}}$dinger-Feynman algorithm to simulating our circuit are put in the Supplemental Material.

\section{Conclusion} \label{sec:summary}
In conclusion, the high-performance superconducting quantum computing system \textit{Zuchongzhi} 2.1 has achieved a larger-scale random quantum circuit sampling with 60-qubit and 24-cycle, yielding a Hilbert space dimension up to $2^{60}$. Compared with \textit{Zuchongzhi} 2.0, our sampling task is about 5000 times more difficult in classical simulation, thus the quantum computational advantage has been significantly enhanced. Our future work would be to investigate the practical applications of random quantum circuits on NISQ devices, such as certified random bits~\cite{aaronson2018certified}, error correction~\cite{gullans2020quantum}, and hydrodynamics simulation~\cite{richter2021simulating}, beyond the abstract sampling task.

\begin{acknowledgments}
We thank Run-Ze Liu, Wen Liu, Chenggang Zhou for very helpful discussions and assistance. The classical calculations were performed on the supercomputing system in the Supercomputing Center of University of Science and Technology of China. The design and quality check of the sample are completed using SQCEDA packages (https://github.com/sqceda). The authors thank the USTC Center for Micro- and Nanoscale Research and Fabrication for supporting the sample fabrication. The authors also thank QuantumCTek Co., Ltd., for supporting the fabrication and the maintenance of room-temperature electronics.  \textbf{Funding:}
 This research was supported by the National Key R\&D Program of China (Grant No. 2017YFA0304300), the Chinese Academy of Sciences, Anhui Initiative in Quantum Information Technologies, Technology Committee of Shanghai Municipality, National Natural Science Foundation of China (Grants No. 11905217, No. 11774326, Grants No. 11905294), Shanghai Municipal Science and Technology Major Project (Grant No. 2019SHZDZX01), Natural Science Foundation of Shanghai (Grant No. 19ZR1462700), Key-Area Research and Development Program of Guangdong Provice (Grant No. 2020B0303030001), and the Youth Talent Lifting Project (Grant No. 2020-JCJQ-QT-030).
 
 The authors' names appear in alphabetical order by last name.
\end{acknowledgments}

\bibliographystyle{apsrev4-1}
\bibliography{references}

\end{document}

\title{Supplemental Material for \\ ``Quantum Computational Advantage via 60-Qubit 24-Cycle Random Circuit Sampling''}

\author{Qingling Zhu}
\affiliation{Hefei National Laboratory for Physical Sciences at the Microscale and Department of Modern Physics, University of Science and Technology of China, Hefei 230026, China}
\affiliation{Shanghai Branch, CAS Center for Excellence in Quantum Information and Quantum Physics, University of Science and Technology of China, Shanghai 201315, China}
\affiliation{Shanghai Research Center for Quantum Sciences, Shanghai 201315, China}
\author{Sirui Cao}
\author{Fusheng Chen}
\author{Ming-Cheng Chen}
\affiliation{Hefei National Laboratory for Physical Sciences at the Microscale and Department of Modern Physics, University of Science and Technology of China, Hefei 230026, China}
\affiliation{Shanghai Branch, CAS Center for Excellence in Quantum Information and Quantum Physics, University of Science and Technology of China, Shanghai 201315, China}
\affiliation{Shanghai Research Center for Quantum Sciences, Shanghai 201315, China}
\author{Xiawei Chen}
\affiliation{Shanghai Branch, CAS Center for Excellence in Quantum Information and Quantum Physics, University of Science and Technology of China, Shanghai 201315, China}
\author{Tung-Hsun Chung}
\author{Hui Deng}
\affiliation{Hefei National Laboratory for Physical Sciences at the Microscale and Department of Modern Physics, University of Science and Technology of China, Hefei 230026, China}
\affiliation{Shanghai Branch, CAS Center for Excellence in Quantum Information and Quantum Physics, University of Science and Technology of China, Shanghai 201315, China}
\affiliation{Shanghai Research Center for Quantum Sciences, Shanghai 201315, China}
\author{Yajie Du}
\affiliation{Shanghai Branch, CAS Center for Excellence in Quantum Information and Quantum Physics, University of Science and Technology of China, Shanghai 201315, China}
\author{Daojin Fan}
\author{Ming Gong}
\author{Cheng Guo}
\author{Chu Guo}
\author{Shaojun Guo}
\author{Lianchen Han}
\affiliation{Hefei National Laboratory for Physical Sciences at the Microscale and Department of Modern Physics, University of Science and Technology of China, Hefei 230026, China}
\affiliation{Shanghai Branch, CAS Center for Excellence in Quantum Information and Quantum Physics, University of Science and Technology of China, Shanghai 201315, China}
\affiliation{Shanghai Research Center for Quantum Sciences, Shanghai 201315, China}
\author{Linyin Hong}
\affiliation{QuantumCTek Co., Ltd., Hefei 230026, China}
\author{He-Liang Huang}
\affiliation{Hefei National Laboratory for Physical Sciences at the Microscale and Department of Modern Physics, University of Science and Technology of China, Hefei 230026, China}
\affiliation{Shanghai Branch, CAS Center for Excellence in Quantum Information and Quantum Physics, University of Science and Technology of China, Shanghai 201315, China}
\affiliation{Shanghai Research Center for Quantum Sciences, Shanghai 201315, China}
\affiliation{Henan Key Laboratory of Quantum Information and Cryptography, Zhengzhou 450000, China}
\author{Yong-Heng Huo}
\affiliation{Hefei National Laboratory for Physical Sciences at the Microscale and Department of Modern Physics, University of Science and Technology of China, Hefei 230026, China}
\affiliation{Shanghai Branch, CAS Center for Excellence in Quantum Information and Quantum Physics, University of Science and Technology of China, Shanghai 201315, China}
\affiliation{Shanghai Research Center for Quantum Sciences, Shanghai 201315, China}
\author{Liping Li}
\affiliation{Shanghai Branch, CAS Center for Excellence in Quantum Information and Quantum Physics, University of Science and Technology of China, Shanghai 201315, China}
\author{Na Li}
\author{Shaowei Li}
\author{Yuan Li}
\author{Futian Liang}
\affiliation{Hefei National Laboratory for Physical Sciences at the Microscale and Department of Modern Physics, University of Science and Technology of China, Hefei 230026, China}
\affiliation{Shanghai Branch, CAS Center for Excellence in Quantum Information and Quantum Physics, University of Science and Technology of China, Shanghai 201315, China}
\affiliation{Shanghai Research Center for Quantum Sciences, Shanghai 201315, China}
\author{Chun Lin}
\affiliation{Shanghai Institute of Technical Physics, Chinese Academy of Sciences, Shanghai 200083, China}
\author{Jin Lin}
\author{Haoran Qian}
\affiliation{Hefei National Laboratory for Physical Sciences at the Microscale and Department of Modern Physics, University of Science and Technology of China, Hefei 230026, China}
\affiliation{Shanghai Branch, CAS Center for Excellence in Quantum Information and Quantum Physics, University of Science and Technology of China, Shanghai 201315, China}
\affiliation{Shanghai Research Center for Quantum Sciences, Shanghai 201315, China}
\author{Dan Qiao}
\affiliation{Shanghai Branch, CAS Center for Excellence in Quantum Information and Quantum Physics, University of Science and Technology of China, Shanghai 201315, China}
\author{Hao Rong}
\author{Hong Su}
\author{Lihua Sun}
\affiliation{Hefei National Laboratory for Physical Sciences at the Microscale and Department of Modern Physics, University of Science and Technology of China, Hefei 230026, China}
\affiliation{Shanghai Branch, CAS Center for Excellence in Quantum Information and Quantum Physics, University of Science and Technology of China, Shanghai 201315, China}
\affiliation{Shanghai Research Center for Quantum Sciences, Shanghai 201315, China}
\author{Liangyuan Wang}
\affiliation{Shanghai Branch, CAS Center for Excellence in Quantum Information and Quantum Physics, University of Science and Technology of China, Shanghai 201315, China}
\author{Shiyu Wang}
\author{Dachao Wu}
\author{Yulin Wu}
\author{Yu Xu}
\affiliation{Hefei National Laboratory for Physical Sciences at the Microscale and Department of Modern Physics, University of Science and Technology of China, Hefei 230026, China}
\affiliation{Shanghai Branch, CAS Center for Excellence in Quantum Information and Quantum Physics, University of Science and Technology of China, Shanghai 201315, China}
\affiliation{Shanghai Research Center for Quantum Sciences, Shanghai 201315, China}
\author{Kai Yan}
\affiliation{Shanghai Branch, CAS Center for Excellence in Quantum Information and Quantum Physics, University of Science and Technology of China, Shanghai 201315, China}
\author{Weifeng Yang}
\affiliation{QuantumCTek Co., Ltd., Hefei 230026, China}
\author{Yang Yang}
\affiliation{Shanghai Branch, CAS Center for Excellence in Quantum Information and Quantum Physics, University of Science and Technology of China, Shanghai 201315, China}
\author{Yangsen Ye}
\affiliation{Hefei National Laboratory for Physical Sciences at the Microscale and Department of Modern Physics, University of Science and Technology of China, Hefei 230026, China}
\affiliation{Shanghai Branch, CAS Center for Excellence in Quantum Information and Quantum Physics, University of Science and Technology of China, Shanghai 201315, China}
\affiliation{Shanghai Research Center for Quantum Sciences, Shanghai 201315, China}
\author{Jianghan Yin}
\affiliation{Shanghai Branch, CAS Center for Excellence in Quantum Information and Quantum Physics, University of Science and Technology of China, Shanghai 201315, China}
\author{Chong Ying}
\author{Jiale Yu}
\author{Chen Zha}
\author{Cha Zhang}
\affiliation{Hefei National Laboratory for Physical Sciences at the Microscale and Department of Modern Physics, University of Science and Technology of China, Hefei 230026, China}
\affiliation{Shanghai Branch, CAS Center for Excellence in Quantum Information and Quantum Physics, University of Science and Technology of China, Shanghai 201315, China}
\affiliation{Shanghai Research Center for Quantum Sciences, Shanghai 201315, China}
\author{Haibin Zhang}
\affiliation{Shanghai Branch, CAS Center for Excellence in Quantum Information and Quantum Physics, University of Science and Technology of China, Shanghai 201315, China}
\author{Kaili Zhang}
\author{Yiming Zhang}
\affiliation{Hefei National Laboratory for Physical Sciences at the Microscale and Department of Modern Physics, University of Science and Technology of China, Hefei 230026, China}
\affiliation{Shanghai Branch, CAS Center for Excellence in Quantum Information and Quantum Physics, University of Science and Technology of China, Shanghai 201315, China}
\affiliation{Shanghai Research Center for Quantum Sciences, Shanghai 201315, China}
\author{Han Zhao}
\affiliation{Shanghai Branch, CAS Center for Excellence in Quantum Information and Quantum Physics, University of Science and Technology of China, Shanghai 201315, China}
\author{Youwei Zhao}
\affiliation{Hefei National Laboratory for Physical Sciences at the Microscale and Department of Modern Physics, University of Science and Technology of China, Hefei 230026, China}
\affiliation{Shanghai Branch, CAS Center for Excellence in Quantum Information and Quantum Physics, University of Science and Technology of China, Shanghai 201315, China}
\affiliation{Shanghai Research Center for Quantum Sciences, Shanghai 201315, China}
\author{Liang Zhou}
\affiliation{QuantumCTek Co., Ltd., Hefei 230026, China}
\author{Chao-Yang Lu}
\author{Cheng-Zhi Peng}
\author{Xiaobo Zhu}
\author{Jian-Wei Pan}

\affiliation{Hefei National Laboratory for Physical Sciences at the Microscale and Department of Modern Physics, University of Science and Technology of China, Hefei 230026, China}
\affiliation{Shanghai Branch, CAS Center for Excellence in Quantum Information and Quantum Physics, University of Science and Technology of China, Shanghai 201315, China}
\affiliation{Shanghai Research Center for Quantum Sciences, Shanghai 201315, China}

\date{\today}

\maketitle

\setcounter{section}{0}
\renewcommand{\thefigure}{S\arabic{figure}}	
\renewcommand{\thetable}{S\arabic{table}}	
\renewcommand{\theequation}{S\arabic{equation}}	
\setcounter{figure}{0}
\setcounter{table}{0}
\setcounter{equation}{0}

Since most of the methods of calibration and result analysis have been provided in the Ref.~\cite{wu2021strong} of \textit{Zuchongzhi} 2.0, in this supplemental material, we will list the relevant results of \textit{Zuchongzhi} 2.1 without describing too many details about the method. Interested readers may refer to the Ref.~\cite{wu2021strong} for more details about the method used.

\section{Quantum processor}
Our device is a 66-qubit superconducting programmable quantum processor, consisting of $66$ Transmon qubits~\cite{koch:042319} and $110$ tunable couplers~\cite{yysCouplerCZ}, which is similar to \textit{Zuchongzhi} 2.0 ~\cite{wu2021strong}. Each qubit is inductively coupled to a control line for both XY and Z control, and capacitively coupled to a separate $\lambda/4$ readout resonator which is coupled to a $\lambda/2$ bandpass filter~\cite{PhysRevLett.112.190504, zhu2021observation} for fast and high fidelity readout. Each qubit is tunably coupled to one of its nearest neighbor qubits with a coupler which can tune coupling strength from $\sim +5 \text{MHz}$ to $\sim -50 \text{MHz}$, except for the coupler between qubit $Q_{22}$ and $Q_{15}$ which does not work well. With the tunable coupler, we can tune off the qubit-qubit coupling when qubit at idle, and tune on the coupling when performing two-qubit gates. The coupling strength we used to perform two-qubit gates is about $-14 \text{MHz}$. Steps of fabrication are same with \textit{Zuchongzhi} 2.0, which will not be described here.

\section{Experimental wiring}
%接线一模一样，引用一下 ,ADC,DAC,mw
 The experimental wiring setup for qubit$/$coupler control and frequency-multiplexed readout of cryogenic system and the configuration of room temperature electronics are same with the setup of \textit{Zuchongzhi} 2.0~\cite{wu2021strong}. The room temperature electronics used in this experiment includes 330 DAC channels, 11 ADC channels, 11 DC channels and 32 microwave source channels. All experiments are performed in a dilution refrigerator at base temperature of 20mK with a magnetic field shield.

\section{Gate performance and Readout}
\subsection{Basic calibration on 66 qubits}
%与祖冲之一样，经过基本的校准，得到66个比特的二维平均T1，图2DT1，其中G2215不工作。

The basic calibration steps of the processor are also the same with \textit{Zuchongzhi} 2.0, all 66 qubits, 110 couplers, 66 readout resonators and 11 JPAs are involved in this process. One can see more details about the procedure in Ref~\cite{wu2021strong}. The maximum frequencies $f_{01}^{Max}$ of the all 66 qubits are shown in Fig.~\ref{fig.figS1-66bit}.
The coupler between qubit $Q_{22}$ and $Q_{15}$ does not work properly, so there are 109 couplers which we can use to tune the coupling strength between near-neighbor qubtis in this processor.

\subsection{Fine-tune on 60 qubits}

%Considering the crosstalk between XY control lines and the residual coupling strength,
For technical and theory reasons, we select $60$ qubits in the device and set other qubits' frequencies far away from these qubits to avoid residual coupling. We then perform accurate calibration on these 60 qubits. After fine measurement of $T_1$ and XY crosstalk, we preliminary set a suitable idle frequency distribution. However, we find that the $T_1$ performance at idle frequency may change significantly during the experiment. Thus, we quickly measure these qubits' $T_1$ and $T_2^*$ near idle frequencies in parallel and then adjust the idle frequency distribution if the coherence performance of qubits decreases.
%考虑到XYcrosstalk以及残余耦合等，我们选用其中的60个比特进行仔细的校准
%idle,T1,T2*,
\subsubsection{Readout Calibration}
%读取用了二态读取, 说明在idle点读的，全0全1并行标定和随机01标定结果对比,统计分布图
All selected qubits are placed at idle frequencies during measurement. Readout fidelity of each qubit is measured after qubits are prepared at $|00,...,0\rangle$ or $|11,...,1\rangle$. To reduce the influence of energy relaxation and enhance the effective lifetime, we drive the qubits into $|2\rangle$ from $|1\rangle$ before readout~\cite{Elder_2020}. By applying this technology, we can achieve average readout fidelity of 97.74\%. To evaluate the performance of joint readout, we randomly prepare the qubits states $|s_1,s_2,...,s_{60}\rangle$, where $s_i$ = 0 or 1, and measure these randomly states to obtain the average joint read fidelity, corresponding to 0.2304.

\subsubsection{Single-Qubit Gate Calibration}
\begin{figure}[!htbp]
\begin{center}
\includegraphics[width=0.95\linewidth]{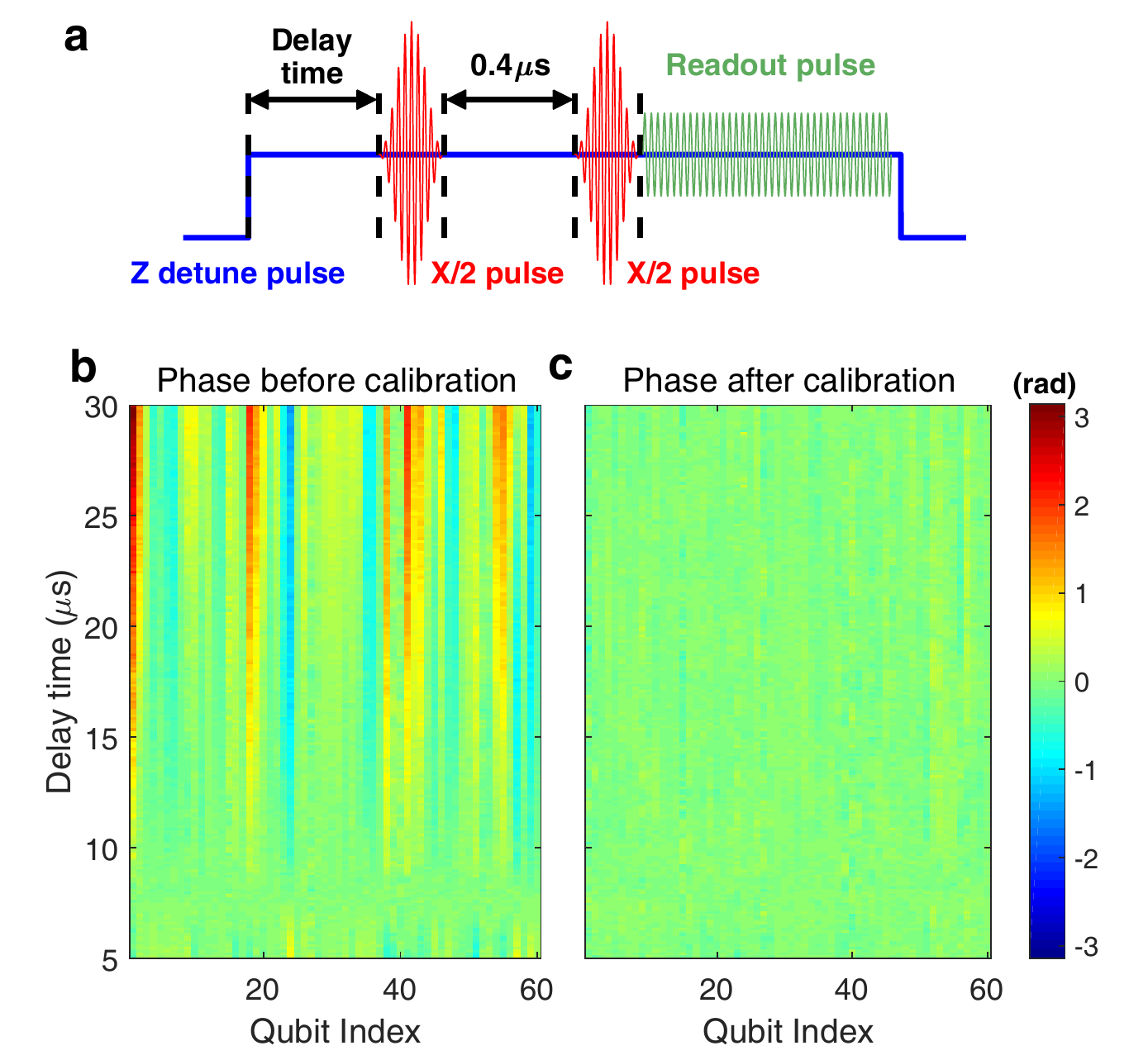}
\end{center}
\caption{\textbf{Z control pulse distortion calibration.} \textbf{a,} Pulse sequence of the z control pulse distortion calibration. \textbf{b,} Phase before calibration. \textbf{c,} Phase after calibration.}
\label{fig.xztail}
\end{figure}

%单比特校准主要将workbias拖尾校准, workbias拖尾校准前后效果对比，单比特门串并行结果，统计分布图
Couplers and qubits are placed at their original frequencies at most time and then detuned to idle frequencies several microseconds before and during the experiment and finally detuned back to their original frequencies after the readout pulses end. Due to this control method, %We use temporary square voltage pulse to tune the frequencies of qubits and couplers. Due to %the distortion of z control pulse,
we find qubit frequency and coupling strength may vary slightly during the operation time, which may reduce the following single-qubit gate fidelity. To mitigate this effect, we measure the qubit frequency variation and then calibrate the square pulse. In Fig. \ref{fig.xztail}, we show our ramsey test pulse to measure z control pulse distortion and the phase before (after) calibration. After this calibration, we measure the single-qubit gate fidelity for isolated operation and simultaneous operation by cross-entropy benchmarking(XEB)~\cite{arute2019quantum}, see Table \ref{tableSummary}. The difference between isolated fidelity and simultaneous fidelity is caused by XY crosstalk and residual coupling strength.

\begin{figure}[!htbp]
\begin{center}
\includegraphics[width=0.95\linewidth]{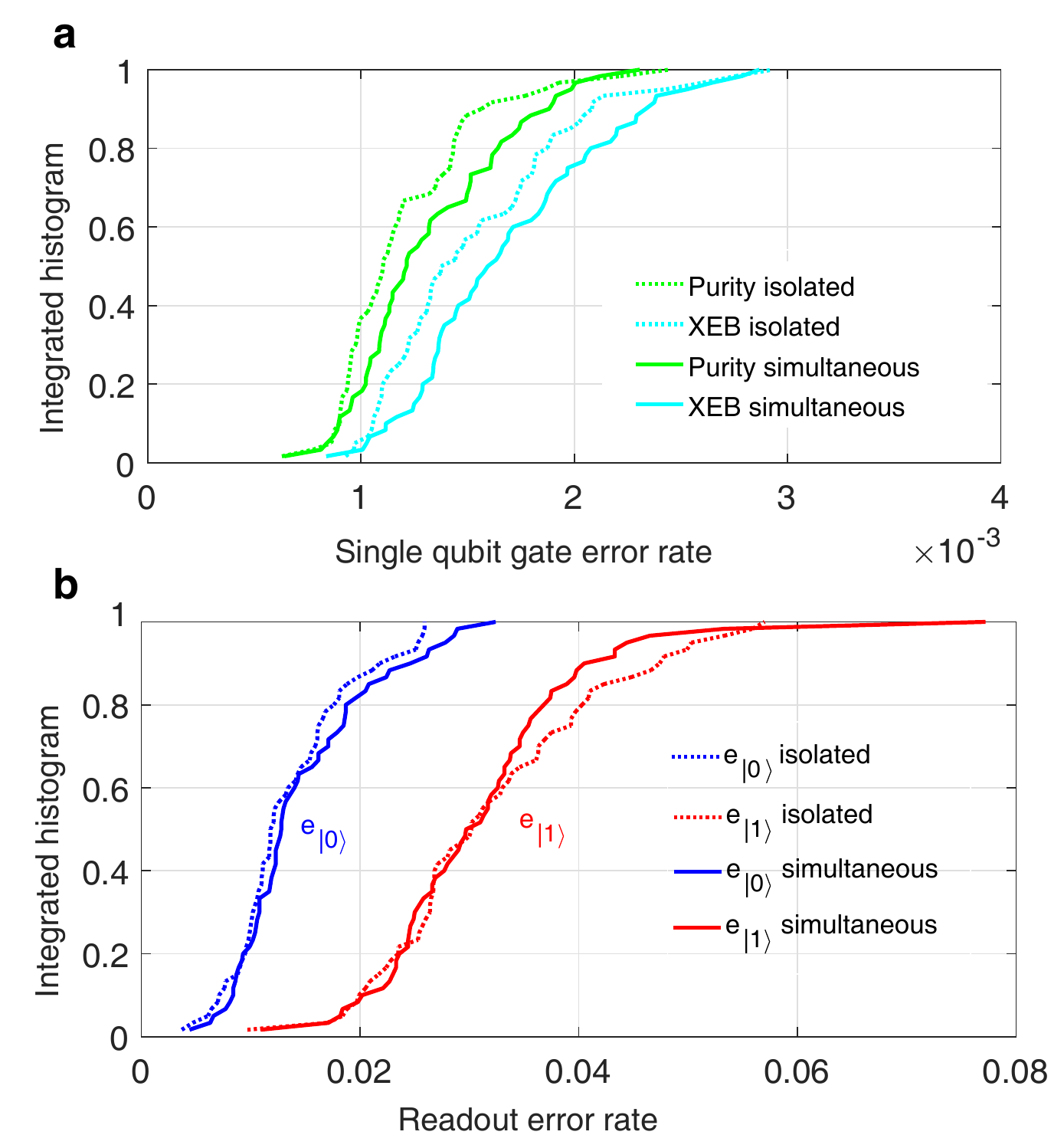}
\end{center}
\caption{\textbf{Single-qubit gates XEB and readout errors.}\textbf{ a,} Integrated histograms of single-qubit gates Pauli error $e_1$. The dotted line represent isolated results, and the line represent simultaneous results. \textbf{b,} Integrated histograms of readout errors $e_{|0\rangle}$ and $e_{|1\rangle}$.}
\label{fig.hf_error_1q_population}
\end{figure}

\begin{figure*}[!htbp]
\begin{center}
\includegraphics[width=0.7\linewidth]{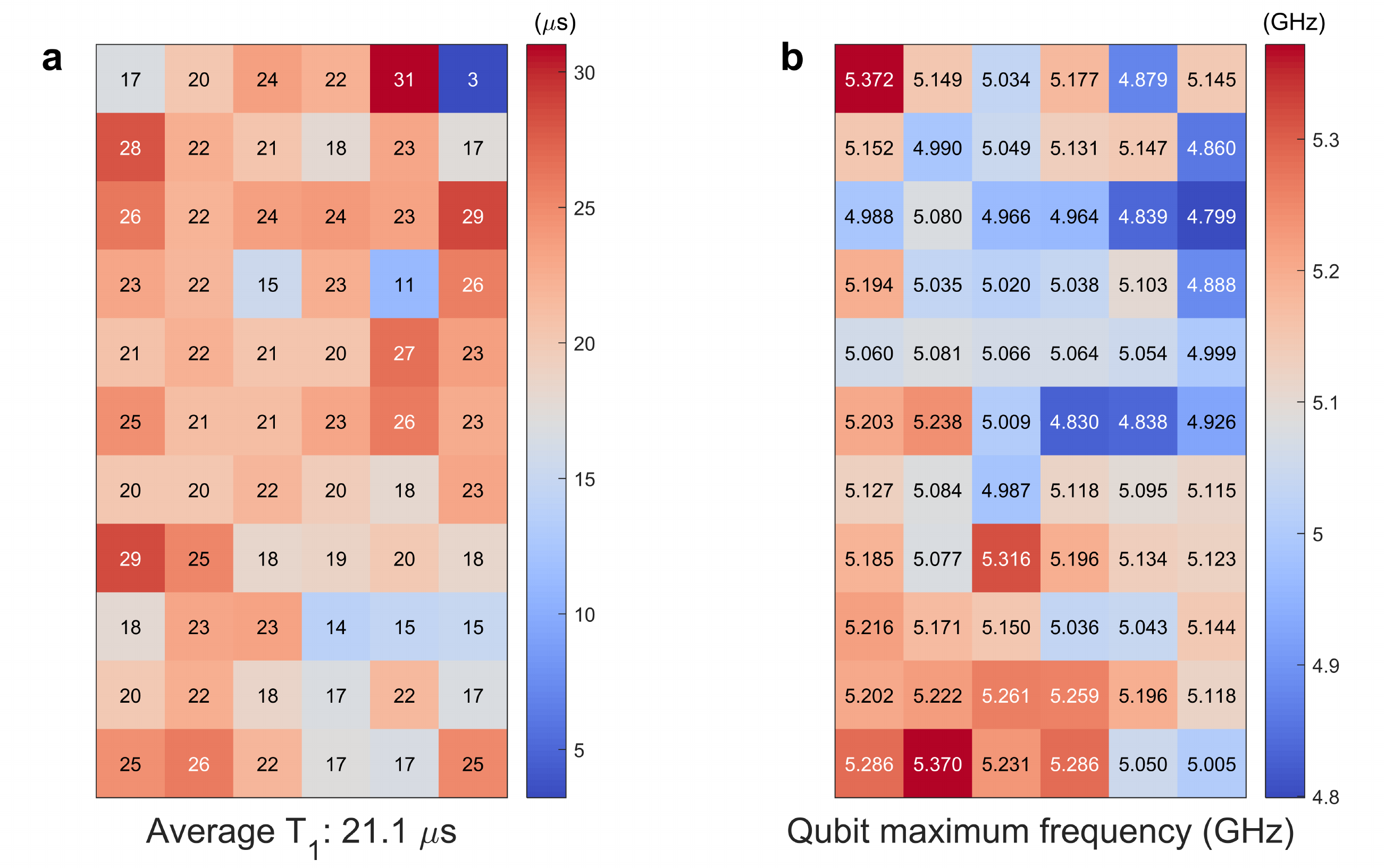}
\end{center}
\caption{\textbf{Average 2D $T_{1}$ and qubit maximum freqeuncy of 66 qubits.}}
\label{fig.figS1-66bit}
\end{figure*}

\begin{figure*}[!htbp]
\begin{center}
\includegraphics[width=0.9\linewidth]{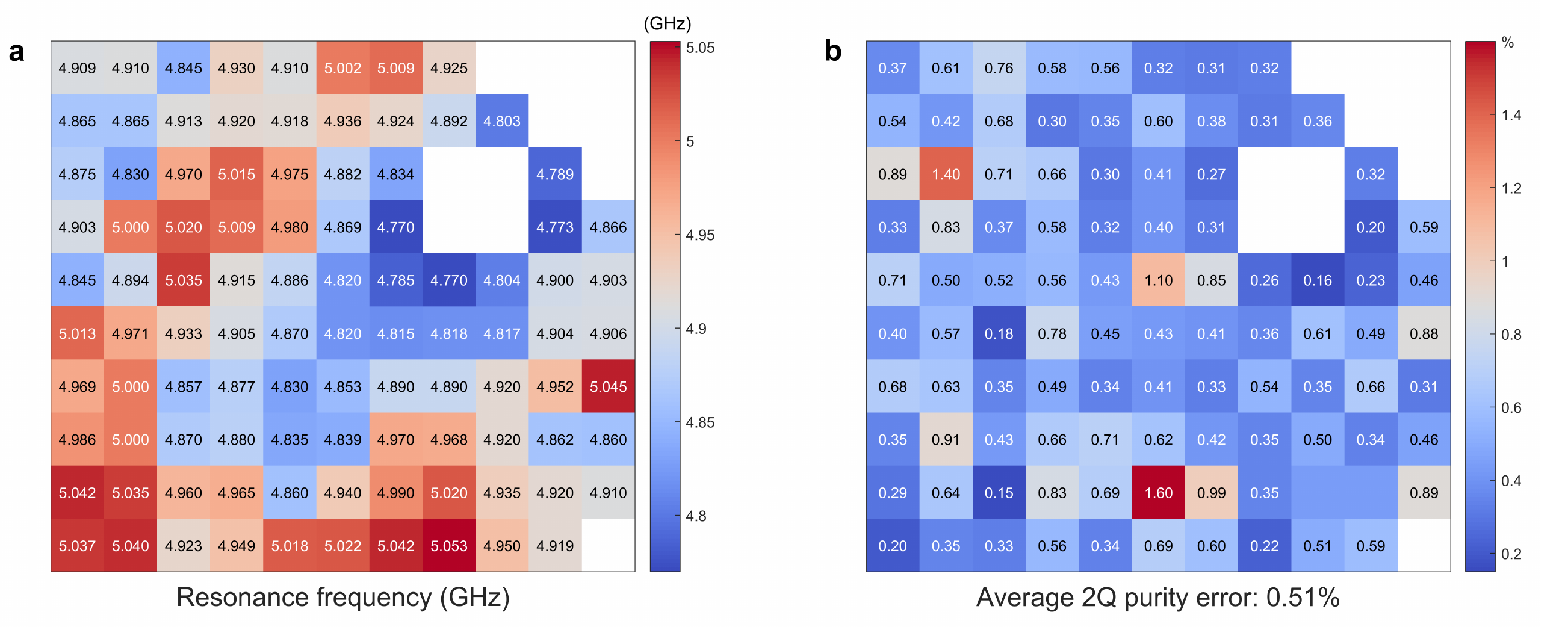}
\end{center}
\caption{\textbf{The qubit resonance frequencies and Purity error distribution of 99 couplers.}
\label{fig.figS3-99coupler}}
\end{figure*}

\begin{figure*}[!htbp]
\begin{center}
\includegraphics[width=0.7\linewidth]{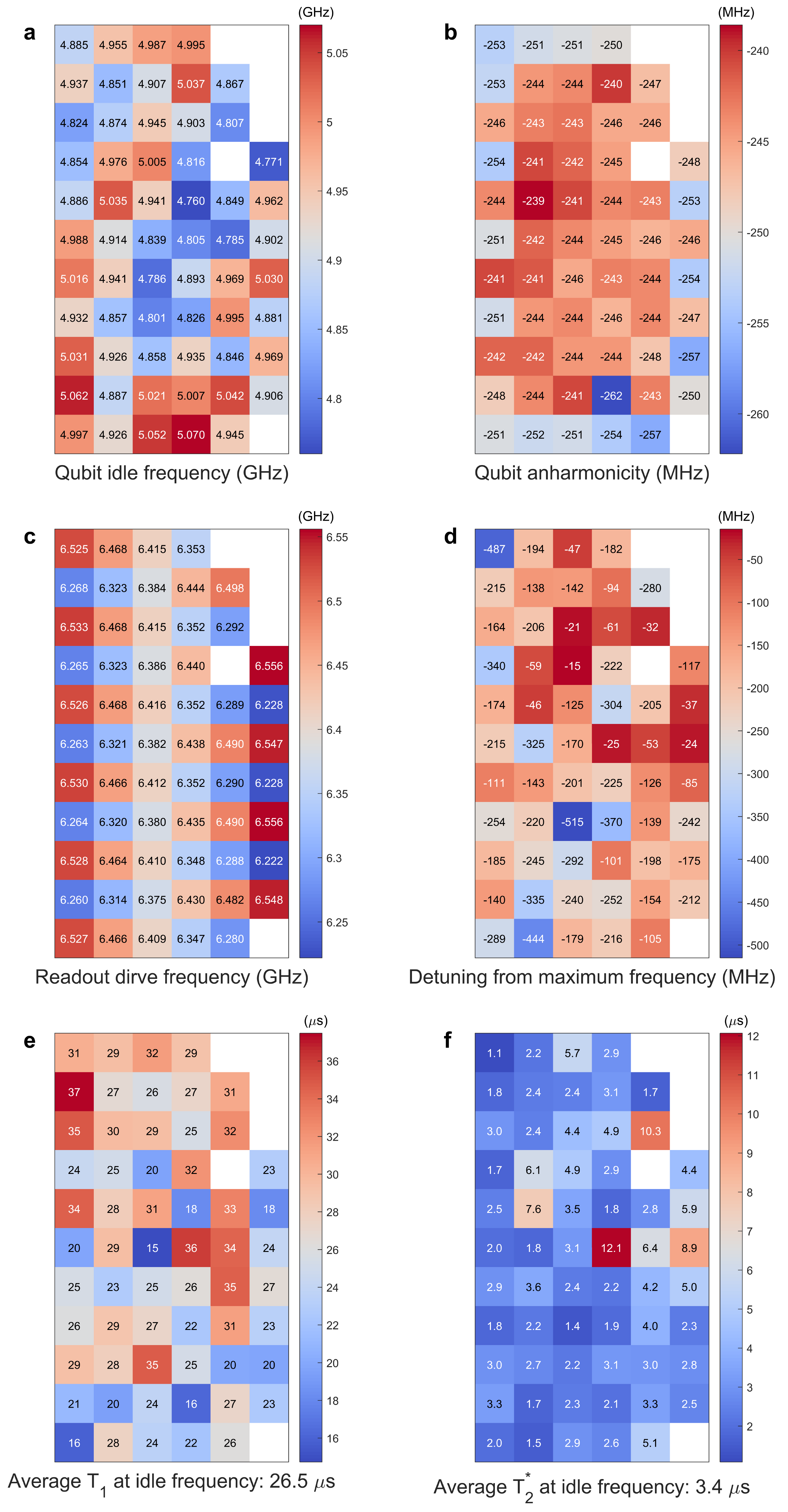}
\end{center}
\caption{ Typical distribution of selected 60-qubits.}
\label{figS2-60bit}
\end{figure*}

\subsubsection{Two-Qubit Gate Calibration}
iSWAP-like gates~\cite{arute2019quantum} are used in our experiment, which are realized by tuning two near-neighbor qubits into resonance. The procedures of calibration and optimizing are similar to \textit{Zuchongzhi} 2.0 which we briefly illustrate here again. First, we set the total duration time of iSWAP-like gates to $24ns$ after considering the effect of the leakage to non-computational basis. To realize swap angles close to $\pi/2$, we delicately calibrate the coupling strength and detuning of two near-neighbor qubits.

\begin{figure}[!htbp]
\begin{center}
\includegraphics[width=0.95\linewidth]{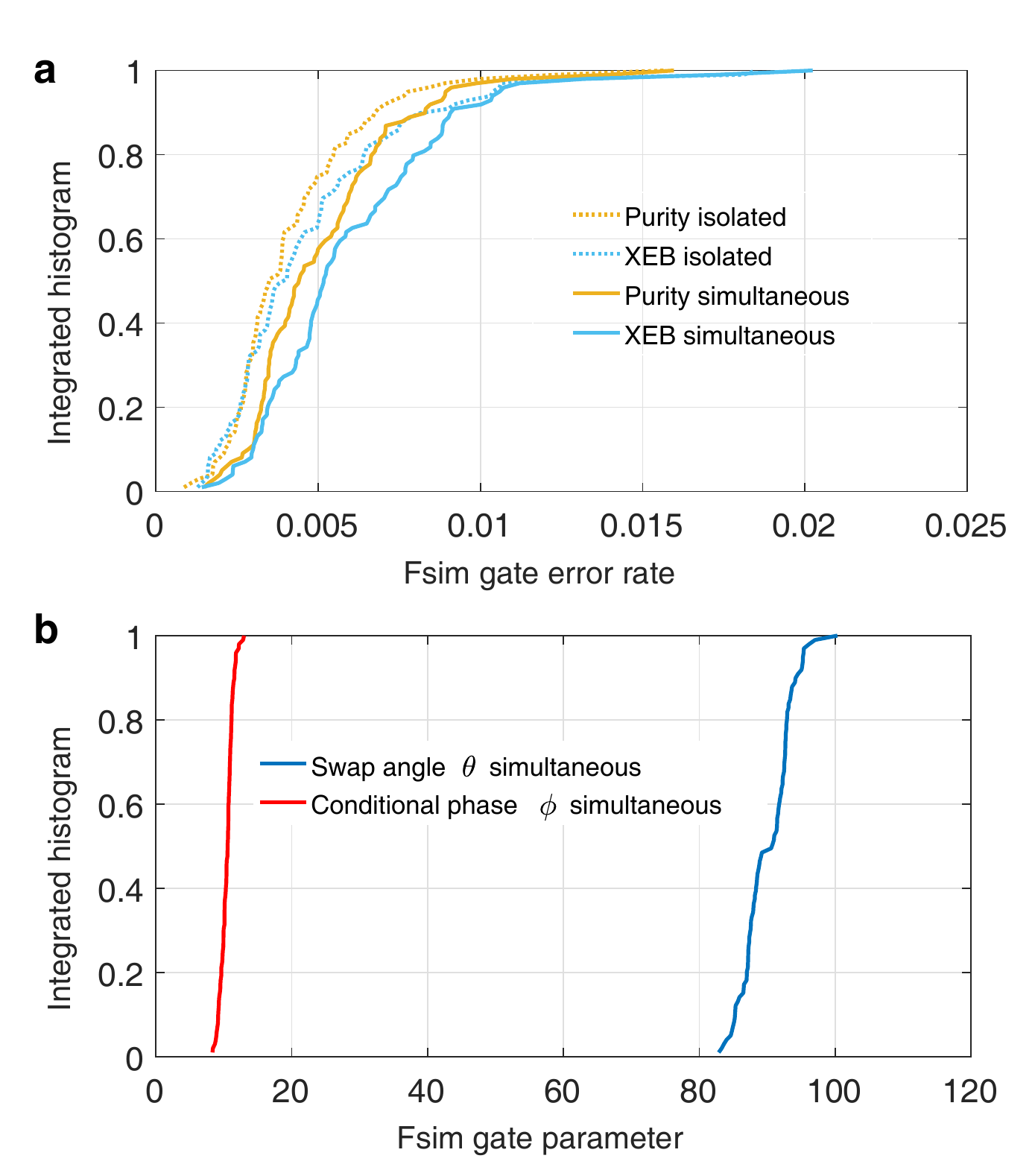}
\end{center}
\caption{\textbf{Two-qubits gates XEB and parameters.}\textbf{ a,} Integrated histograms of two-qubit gates Pauli error $e_2$. The dotted line represent isolated results, and the line represent simultaneous results. \textbf{b,} Integrated histograms of two control parameters $\theta$ and $\phi$.
}
\label{fig.hf_error_2q_population}
\end{figure}

After calibration of iSWAP-like gates parameters, we use XEB~\cite{arute2019quantum} to benchmark the gate performance. Through optimize the five parameters $\{\theta, \phi, \Delta_{+}, \Delta_{-}, \Delta_{-,\text{off}}\}$, as described in the main text, overall fidelities of XEB are obtained Fig.~2(main). We also analyze the error of speckle purity benchmarking(SPB)~\cite{arute2019quantum} which is shown in Fig.~\ref{fig.figS3-99coupler}(b). Considering the effects of two-level-system(TLS), pulse distortion, decoherence and the residual coupling between nearest-neighbor qubits when simultaneous manipulating two-qubit gates in a specific pattern, we obtain an optimized resonance frequency configuration of all $99$ two-qubit gates in four different patterns, as exhibited in Fig.~\ref{fig.figS3-99coupler}(a).

After calibration and optimization, we compare two-qubit gates performance in the isolated and simultaneous situations. As illustrated in Fig.~\ref{fig.hf_error_2q_population}(a), the XEB error is close to SPB error in both cases which means our control error is small. The difference is that the SPB error is larger when simultaneously manipulate two-qubit gates which mainly caused from the residual coupling strength. Partial parameters of all $99$ iSwap-like gates are shown in Fig.~\ref{fig.hf_error_2q_population}(b), the average swap angle and conditional phase are $90$ degrees and $10$ degrees separately.

\begin{table*}[htb!]
    \centering
    \begin{tabular}{p{6cm}p{2.5cm}<{\centering}p{2.5cm}<{\centering}p{2.5cm}<{\centering}p{2.5cm}<{\centering}}
    \toprule
    Parameters& Mean& Median& Stdev.& Figure\\
    \toprule

    Qubit maximum frequency (GHz)& 5.094& 5.099& 0.129& Fig.~\ref{fig.figS1-66bit}\\
    Qubit idle frequency (GHz)& 4.921& 4.926& 0.082& Fig.~\ref{figS2-60bit}\\
    Readout drive frequency (GHz)& 6.397& 6.409& 0.097& Fig.~\ref{figS2-60bit}\\
    Qubit anharmonicity  (MHz)& -246.6& -245.3& 4.9& Fig.~\ref{figS2-60bit}\\
    $T_1$ at idle frequency ($\mu$s)& 26.5& 26.5& 5.3& Fig.~\ref{figS2-60bit}\\
    $T_2^*$ at idle frequency ($\mu$s)& 3.4& 2.9& 2.1& Fig.~\ref{figS2-60bit}\\
    \hline
    Readout $e_{|0\rangle}$ isolated / simultaneous (\%) & 1.36 / 1.48& 1.20 / 1.35& 0.55 / 0.63&  Fig.~\ref{fig.hf_error_1q_population}\\
    Readout $e_{|1\rangle}$ isolated / simultaneous (\%) & 3.22 / 3.03& 3.02 / 2.93& 1.03 / 0.91&  Fig.~\ref{fig.hf_error_1q_population}\\
    1Q XEB $e_{1}$ isolated / simultaneous (\%) & 0.15 / 0.16& 0.14 / 0.16& 0.05 / 0.04&  Fig.~\ref{fig.hf_error_1q_population}\\
    1Q XEB $e_{1}$ purity isolated / simultaneous (\%) & 0.12 / 0.13 & 0.11 / 0.12& 0.03 / 0.04&  Fig.~\ref{fig.hf_error_1q_population}\\
    2Q XEB $e_{2}$ isolated / simultaneous (\%) & 0.48 / 0.60 & 0.40 / 0.52& 0.31 / 0.30&  Fig.~\ref{fig.hf_error_2q_population}\\
    2Q XEB $e_{2}$ purity isolated / simultaneous (\%) & 0.42 / 0.51& 0.35 / 0.44& 0.23 / 0.24&  Fig.~\ref{fig.hf_error_2q_population}\\

    \toprule
    \end{tabular}
    \caption{\textbf{Summary of system parameters.}}
    \label{tableSummary}
    \end{table*}

\subsection{4-patch calibration}
Till now the parameters of all the iSWAP-like gates are calibrated using two-qubit XEB. As stated in the main text and section III (B2), the distortion of the square pulse will cause qubit frequency and coupling strength continuously change slightly during the operation time. However, the square pulses on coupler are not calibrated. Due to that the iSWAP-like gate parameters obtained by two-qubit XEB is fitted using 10-500 cycles circuits, the iSWAP-like gate parameters is slightly inaccurate for shallow circuits.

To optimize the XEB fidelity for the shallow and full circuits, it may be helpful to directly calibrate the iSWAP-like gate on the whole system level with shallow cycle. For such a purpose, we will minimize the loss function
\begin{align}\label{eq:loss}
\text{Loss}_{\text{XEB}}(\vec{\theta}, b_{\text{train}}) = 1 - F_{\text{XEB}}(\vec{\theta}, b_{\text{train}}).
\end{align}
Here $\vec{\theta}$ is a list of tunable parameters of all the iSWAP-like gates, $b_{\text{train}}$ is a training dataset which is generated experimentally before hand, $F_\text{XEB}$ is the cross-entropy fidelity. We note that Eq.(\ref{eq:loss}) is a pure classical function except that the training data $b_{\text{train}}$ is generated with a quantum computer. Therefore we can compute the gradient of Eq.(\ref{eq:loss}) using the classical auto-differentiation framework, and then we use the BFGS optimizer to minimize the loss function. The initial values of the iSWAP-like gates are set as the parameters obtained by two-qubit XEB. After that we use the optimal set of parameters, $\vec{\theta}^{\text{optimal}}$, as the final parameters for the iSWAP-like gates.

One difficulty of this approach is that for the full circuit of $60$ qubits, $F_{\text{XEB}}$ will be extremely difficult to evaluate. To overcome this problem, we divide the full circuit into $4$ non-overlapping patches (see Fig.~\ref{fig.4_patch} (a)), which nevertheless covers all the iSWAP gates. We then optimize the parameters in each patch independently, with $b_{\text{train}}$ generated for each patch with depth $24$. After that, we obtain $\vec{\theta}^{optimal}$ for all the iSWAP-like gates.

Figure~\ref{fig.4patchparameters} shows the parameters of all the iSWAP-like gates before and after 4-patch calibration. We can see that the  two parameters $\theta$ and $\phi$ related to the coupling strength have slightly changed before and after 4-patch, which is consistent with our inference. As shown in Fig.~\ref{fig.4_patch} (b), the XEB fidelities of cirucit with 15-60 qubits and 10-cycle can be greatly improved after 4-patch calibration.

\begin{figure}[!htbp]
\begin{center}
\includegraphics[width=0.8\linewidth]{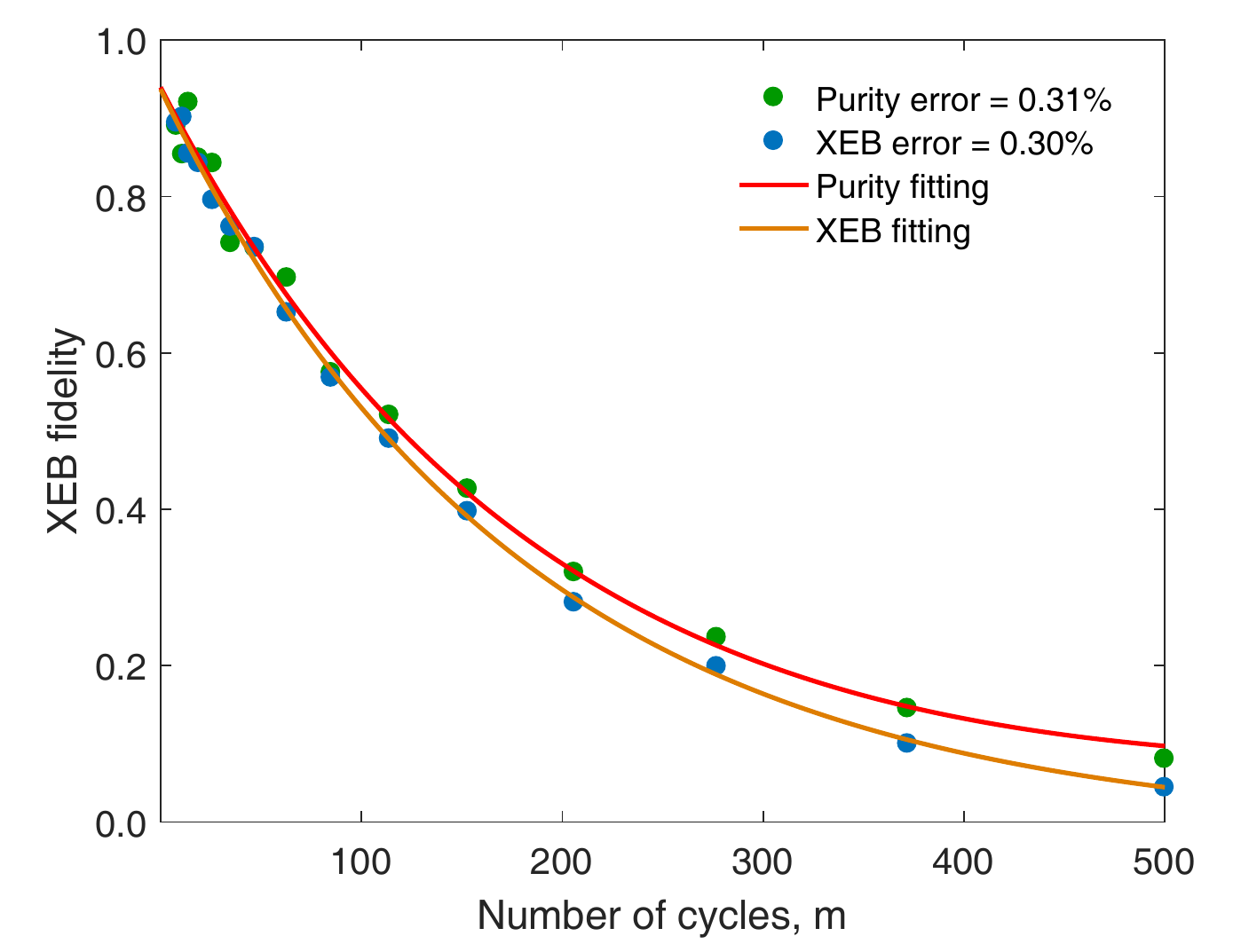}
\end{center}
\caption{\textbf{Two-qubit crossentropy benchmarking (XEB) on a pair of qubits.}}
\label{fig.fsim_gate}
\end{figure}

\begin{figure*}[!htbp]
\begin{center}
\includegraphics[width=1.0\linewidth]{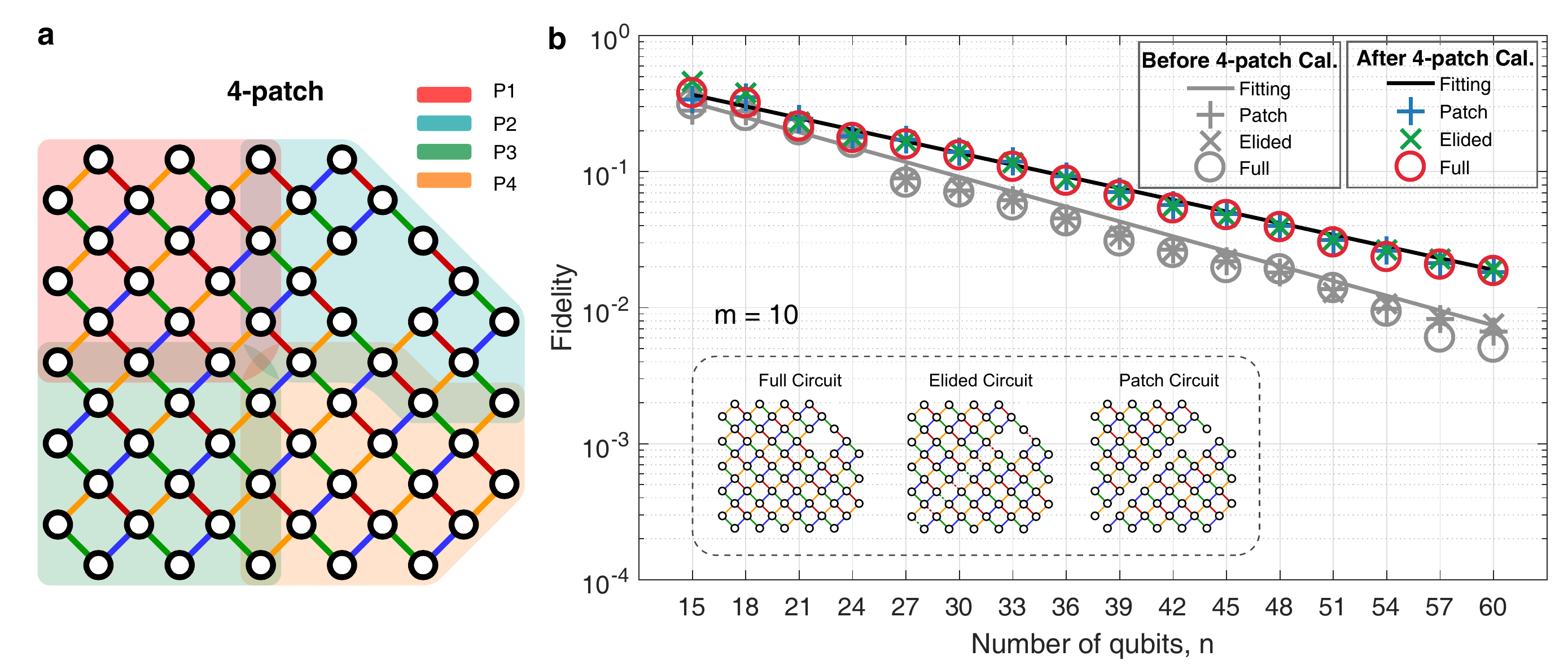}
\end{center}
\caption{\textbf{4-patch calibration and experimental results.} (a) We divide the circuit into 4 patches as shown, and optimize the parameters of iSWAP gates in each patch. (c) The XEB fidelities of cirucit with 15-60 qubits and 10-cycle before and after 4-patch calibration.}
\label{fig.4_patch}
\end{figure*}

\begin{figure}[!htbp]
\begin{center}
\includegraphics[width=0.95\linewidth]{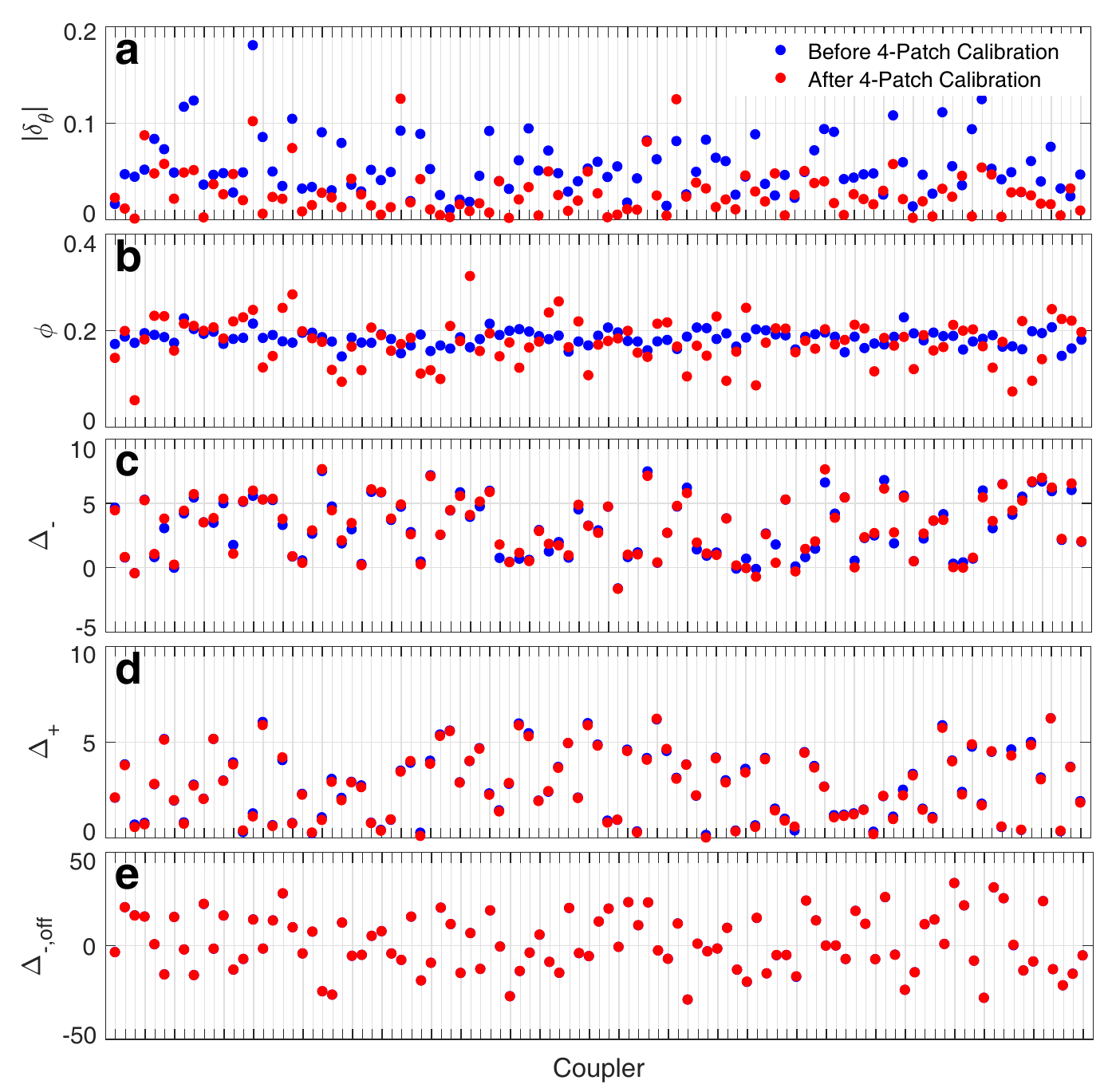}
\end{center}
\caption{\textbf{Parameters of iSWAP-like gates before and after 4-patch calibration}}
\label{fig.4patchparameters}
\end{figure}

\subsection{Summary of system parameters}
The system parameters of our quantum processor 60bit are summarized in Table~\ref{tableSummary}.

\begin{figure*}[!htbp]
\begin{center}
\includegraphics[width=1.0\linewidth]{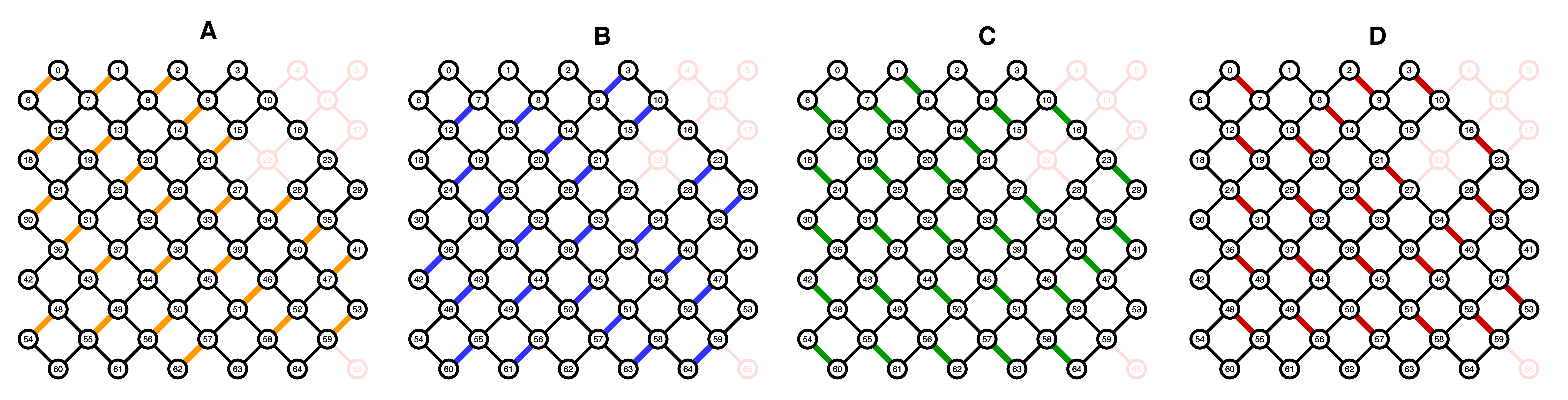}
\end{center}
\caption{\textbf{Coupler activation patterns for 60 qubits.} Coupler activation pattern for 60 qubits that determines which qubits are allowed to interact simultaneously in a cycle.}
\label{fig.ABCD_pattern}
\end{figure*}

\section{Random quantum circuits}

In the random quantum circuit, single-qubit gates are chosen randomly from $\{\sqrt X, \sqrt Y, \sqrt W \}$ , the two-qubit gates are implemented in the sequence of ABCDCDAB, where A, B, C and D are 4 different patterns. The random quantum circuit for 60-qubit and 24-cycle is constructed using the method in Ref.~\cite{wu2021strong}, and the four patterns for two-qubit gates are shown in Fig.~\ref{fig.ABCD_pattern}. For the 60-qubit 22-cycle circuit, the sequence for two-qubit gates are set as (ABCDCDAB)*2+ABCD+CB, to ensure the difficulty of classical simulation.

%(ABCDCDAB)*3, 54 6WEDGE 3DCD 1START 1END
%(ABCDCDAB)*2+ABCD+CB   48 4WEDGE 3DCD 2 START
%and the corresponding promising cut, respectively.

\section{XEB result analysis}

%In theory, circuits of enough depth exhibit the Porter-Thomas distribution.
For a set of bitstrings $\{q_i\}$, We estimate the linear XEB fidelity from the ideal probabilities $\{p_i=p_s(q_i)\}$ as
\begin{equation}
F_l = \langle Dp\rangle  - 1
\label{linearXEB}
\end{equation}

\begin{figure}[htbp]
\begin{center}
\includegraphics[width=0.9\linewidth]{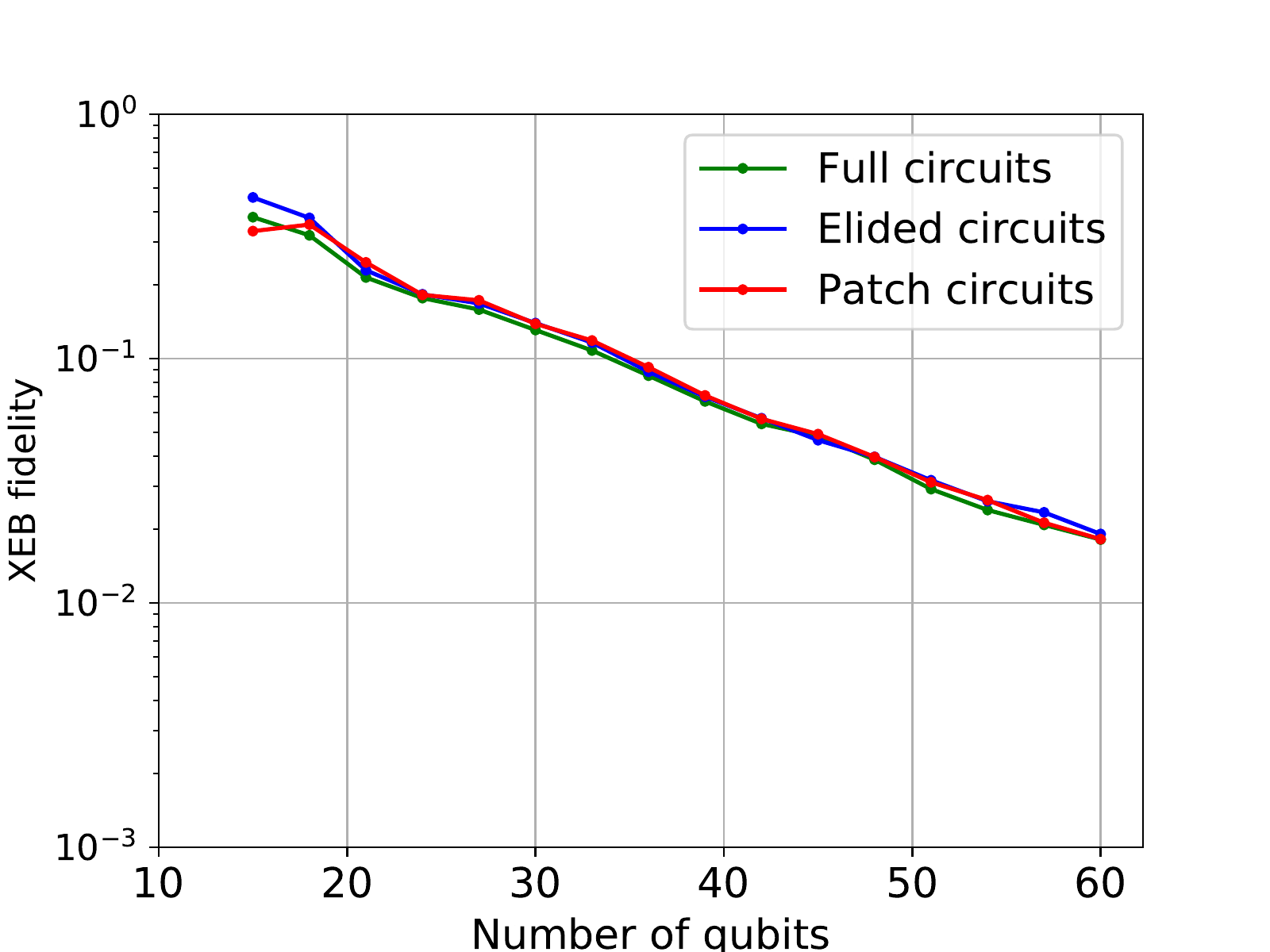}
\end{center}
\caption{\textbf{Performance of patch circuits and elided circuits.} Patch, elided and full circuit XEB fidelity from 15 to 60 qubits with 10 cycles. Each data point is averaged over 9 quantum circuit instances.}
\label{XEB_Patch}
\end{figure}

Firstly, we compare the XEB results of patch circuits and elided circuits with full circuits to check the efficacy of circuit performance estimation method. Fig.~\ref{XEB_Patch} shows the XEB results measured by full, patch and elided circuit of systems from 15 to 60 qubits with 10 cycles. The average ratio of patch circuit and elided circuit fidelity to full circuit fidelity are 1.05 and 1.07, both with a standard deviation of 6\%.

\begin{figure}[!htbp]
\begin{center}
\includegraphics[width=0.9\linewidth]{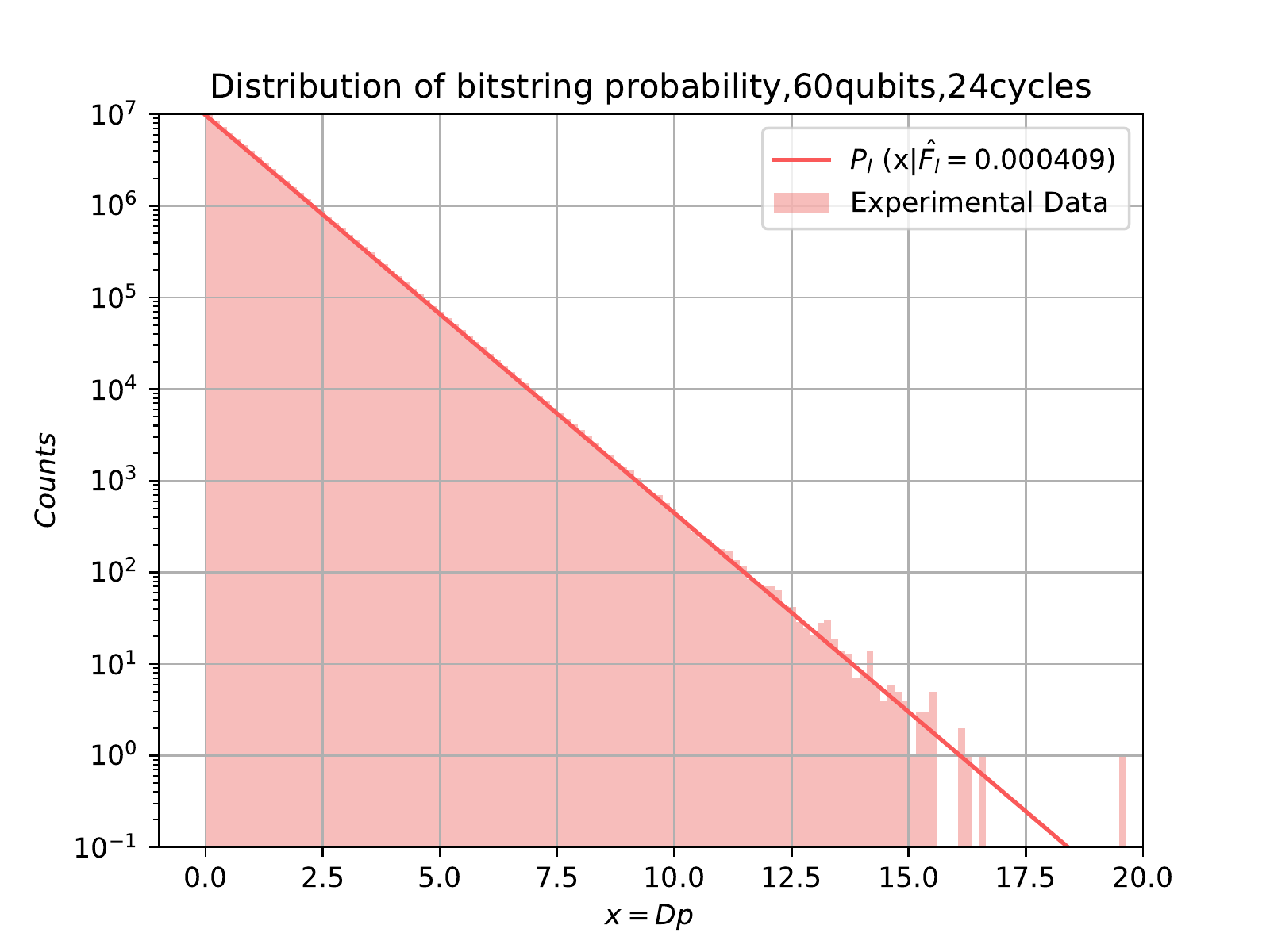}
\end{center}
\caption{\textbf{Distribution of bitstring probabilities from a 60-qubit 24 -cycle circuit.} The theoretical curve $P_l (x|\hat{F_l} )$ is computed with the experimental XEB fidelity and scaled by the number of samples $N_s = 7\times 10^{7}$.}
\label{Fig.1}
\end{figure}

%After sampling the bitstrings, we calculate the ideal probability of each bitstring. Then we examine the distribution of probabilities.
We then check the probability distribution of our samples against the theoretical PDF
\begin{equation}
  P_l (x|\hat{F_l} ) = (\hat{F_l} x+(1-\hat{F_l} ))e^{-x}	
\end{equation}
where $x \equiv Dp$ is bitstring probability scaled by the dimension $D$. For elided circuit with 60 qubits and 24 cycles, we show the result of one circuit instance in Fig.~\ref{Fig.1}. Kolmogorov-Smirnov test is performed to quantify the consistency of experimental data distribution with theoretical curve. Taking one circuit instance for example, the K-S test result is $p = 0.77$ for hypothesis $F=\hat{F}$ while $p=8.0\times 10^{-4}$ for hypothesis $F=0$.

\begin{figure}[tbp]
\begin{center}
\includegraphics[width=0.9\linewidth]{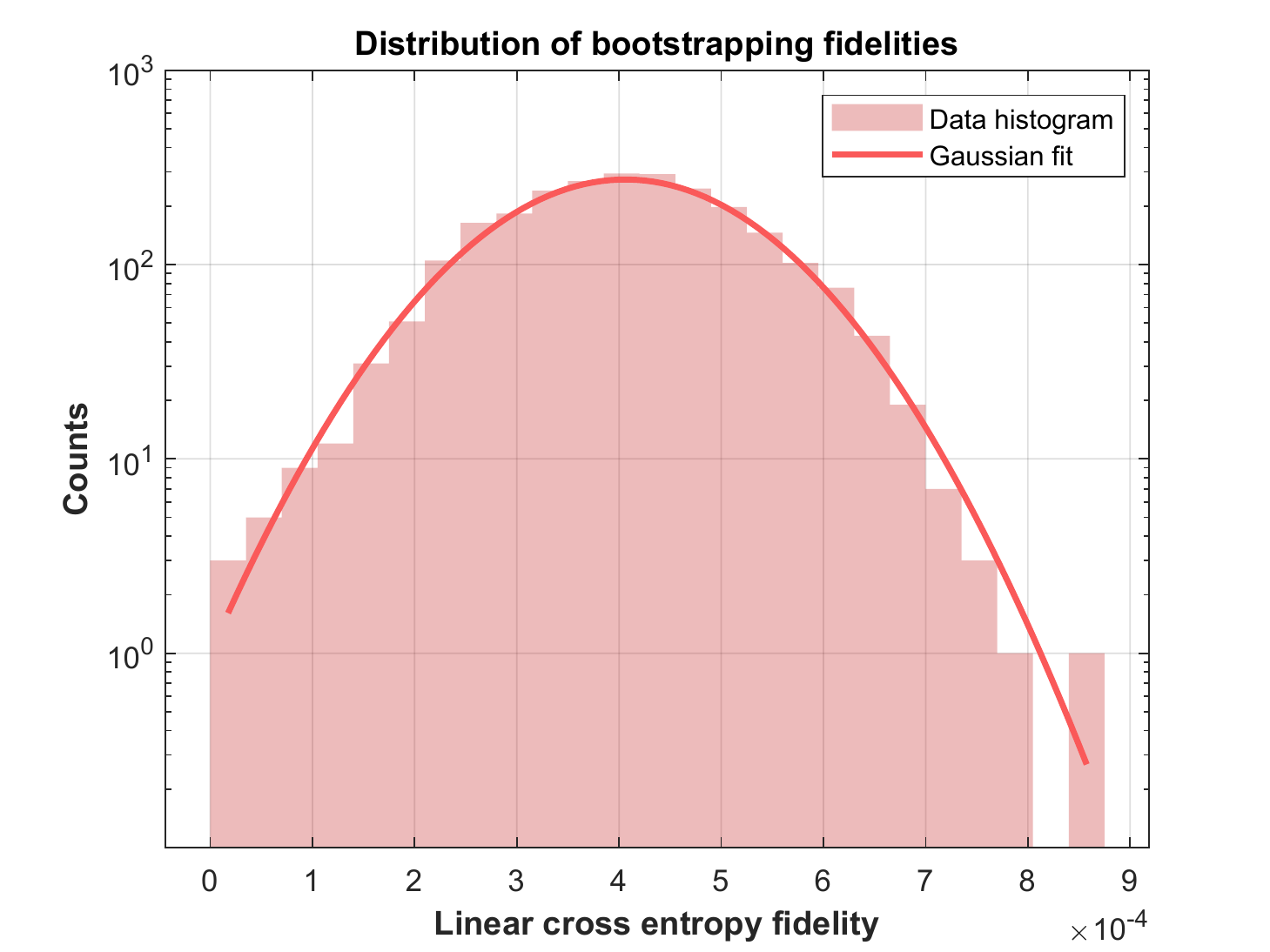}
\end{center}
\caption{\textbf{Fidelity distribution of bootstrap samples.} The samples of a 60-qubit 24-cycle elided circuit are bootstrapped 2500 times. Calculated fidelities distribution is shown with a gaussian fit.}
\label{Fig.3}
\end{figure}

%\subsection{XEB uncertainties}
%The uncertainties of XEB measurements include statistical uncertainties and systematic uncertainties.
The statistical uncertainty of XEB is estimated with formula
\begin{equation}
\hat{\sigma}_{F_{l}}=D \sqrt{\operatorname{Var}(p) / N_{s}}
\label{eq.3}
\end{equation}
We check the formula with bootstrap method. For a 60-qubit 24-cycle elided circuit, the statistical uncertainty of XEB is $1.20, 1.19, 1.21 \times 10^{-4}$ for the esimating formula, bootstrap and gaussian fit. The fidelity distribution of 2500 bootstrap samples with Gaussian fit are shown in Fig.~\ref{Fig.3}.

Then we use inverse-variance weighting to estimate the fidelity and statistical uncertainty of all twelve 60-qubit and 24-cycle elided circuits. The result is $\hat{F_{l}} = (3.66\pm 0.35)\times 10^{-4}$, indicating great consistency with the theoretical statistical uncertainty $3.5 \times 10^{-5}$.

\section{Classical simulation}
In this section, two state-of-the-art classical algorithms, tensor network contraction algorithm (abbreviated as TNC, a single-amplitude algorithm) and Schr$\ddot{\text{o}}$dinger-Feynman algorithm (abbreviated as SFA, a full-amplitude algorithm), are used to  estimate the classical computational cost of our random quantum circuit. Compared with SFA, TNC is more efficient for this system size, so the results of TNC are present in the main text as the estimation of classical computational cost.

\subsection{Tensor network contraction}

\begin{figure}[!htbp]
\begin{center}
\includegraphics[width=\linewidth]{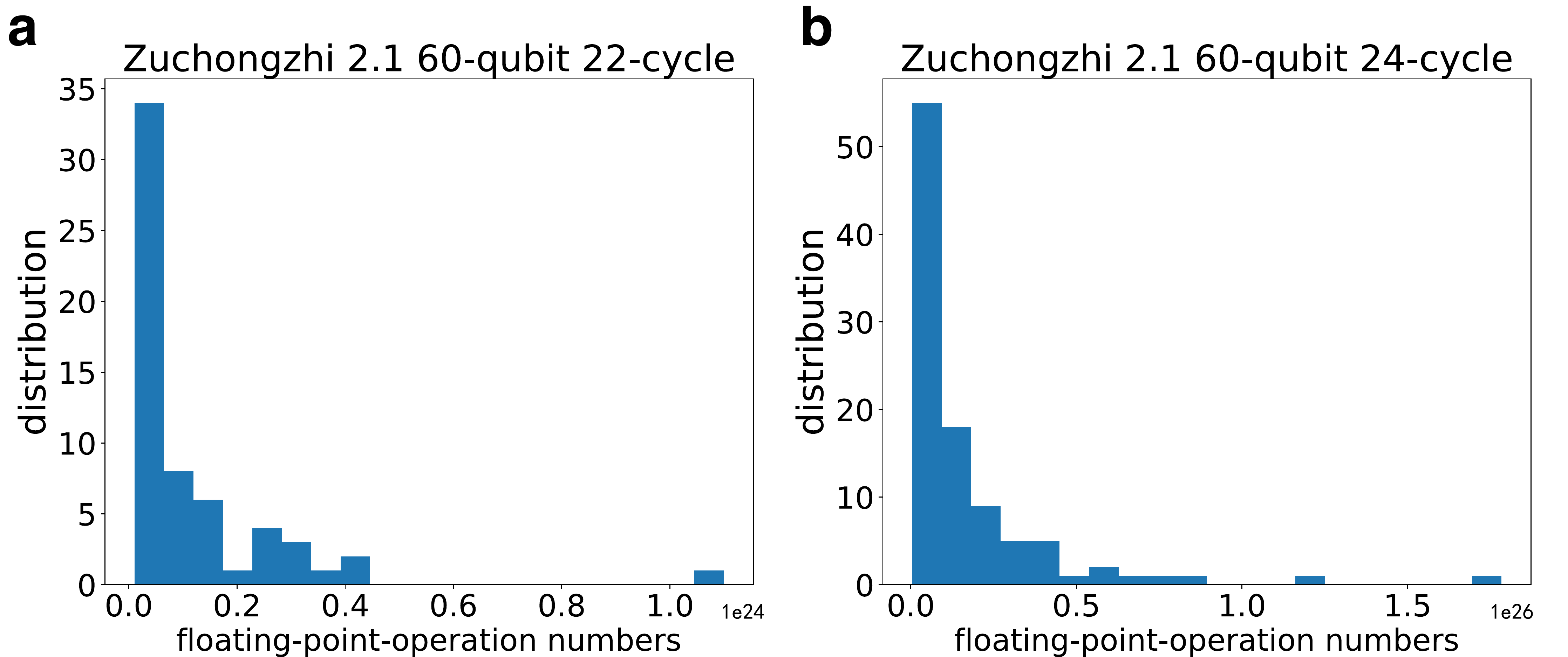}
\end{center}
\caption{\textbf{Probability distributions of the number of floating point operations for contraction of  (a) 60-qubit 22-cycle circuit and (b) 60-qubit 24-cycle circuit.}}
\label{TNCdistribution}
\end{figure}

The cost of TNC algorithm can be estimated by the python package cotengra. When using this package, we leave 21 out of the 60 qubits as open, set the size of the largest intermediate tensor as $2^{30}$, and the optimizer \textit{kahypar} is chosen. For the 60-qubit random quantum circuit with 22 cycles and 24 cycles, the number of floating point operations to generate one perfect sample are estimated as $1.06\times10^{22}$ and $4.68\times10^{23}$, respectively, by repeating the procedure about 100 times (see Fig.~\ref{TNCdistribution}).

As estimated in Ref.~\cite{HuangChen2020}, it would cost 833.75s to generate one perfect sample for a tensor network with $6.66\times 10^{18}$ contraction cost using Summit. Thus, it would cost $\frac{ (4.68 \times {{10}^{23}}) \times (7.0 \times {10^7}) \times (3.66\times 10^{-4})}{{6.66 \times {{10}^{18}}}} \times 833.75\text{s}{\rm{ = }}4.8 \times 10^4$ years to reproduce the same results as the 53-qubit 20-cycle circuit in Ref.~\cite{arute2019quantum} and our 60-qubit 24-cycle circuit using Summit.

\subsection{Schr$\ddot{\text{o}}$dinger-Feynman algorithm}
\begin{figure}[tbp]
\begin{center}
\includegraphics[width=\linewidth]{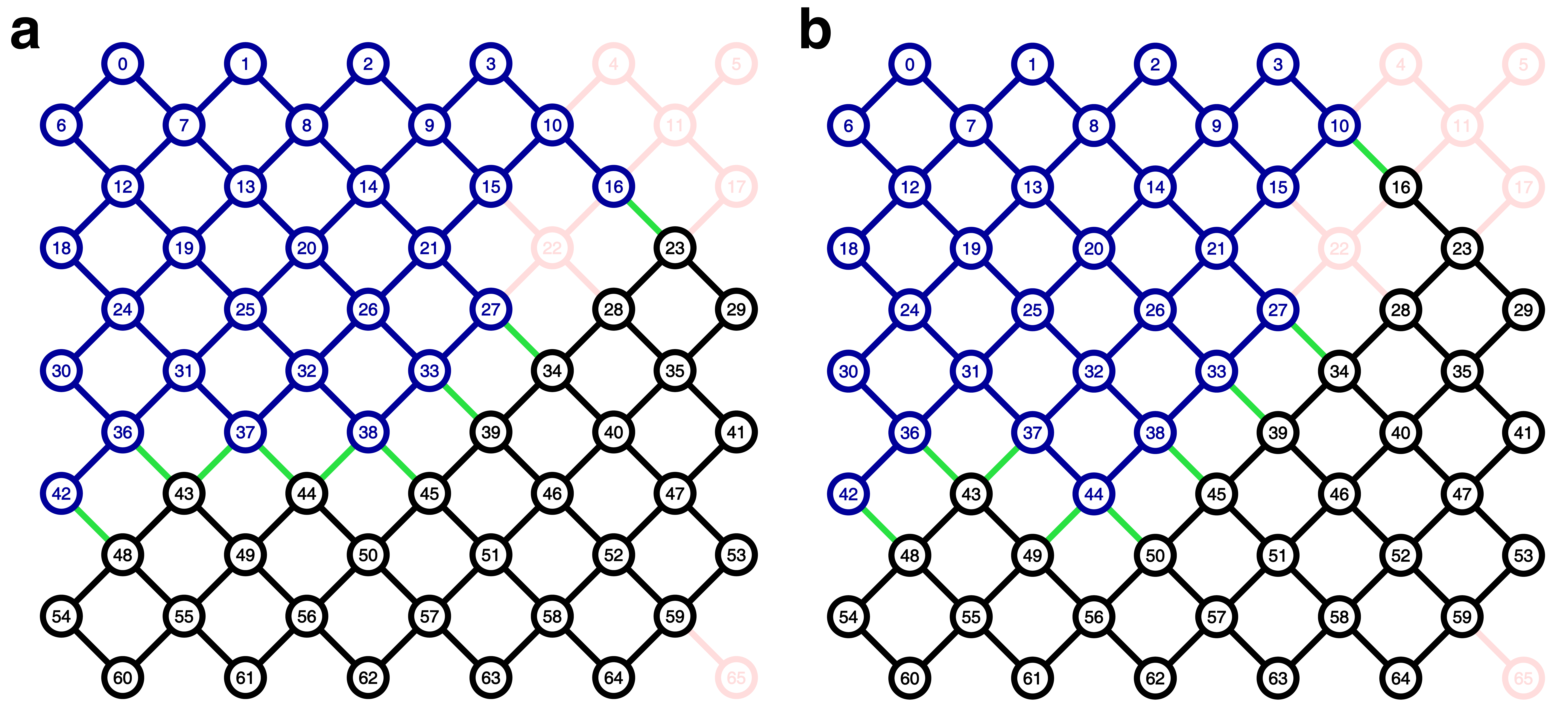}
\end{center}
\caption{\textbf{optimal cut for (a) 60-qubit circuit with 22 cycles  and 60-qubit circuit with 24 cycles, respectively.}.
\label{cut}}
\end{figure}

For the 60-qubit random quantum circuit with 22 cycles and 24 cycles, the promising cuts for SFA are shown in Fig.~\ref{cut}. We use the same simulator and server in Ref.~\cite{wu2021strong} to estimate the cost of SFA. The results are shown in Table~\ref{tab:qsimhruntime} (For the sake of comparison, we also list the estimation for Sycamore and \textit{ZuChongzhi }2.0's circuits).

\begin{table}[htbp]
\centering
\newcommand{\tabincell}[2]{\begin{tabular}{@{}#1@{}}#2\end{tabular}}
\begin{tabular}{|c|c|c|c|c|}
\hline
Processor & \# of qubits & cycle & fidelity & run time (years) \\
\hline
Sycamore & $53 $ & 20    & 0.224$\%$ &  1,332  \\
\textit{Zuchongzhi} 2.0 &$56 $ & 20  & 0.0662$\%$ &  8,612,623  \\
\textit{Zuchongzhi} 2.1 &$60 $ & 22  & 0.0758$\%$ &  $6.4\times 10^{9}$ \\
\textit{Zuchongzhi} 2.1 &$60 $ & 24  & 0.0366$\%$ &  $1.2\times 10^{12}$  \\
\hline
\end{tabular}
\caption{\textbf{Run times of SFA using 7,630,848 CPU cores} (the most powerful supercomputer Fugaku has a total of 7,630,848 cores).  \label{tab:qsimhruntime} }
\end{table}

\begin{figure}[htbp]
\begin{center}
\includegraphics[width=\linewidth]{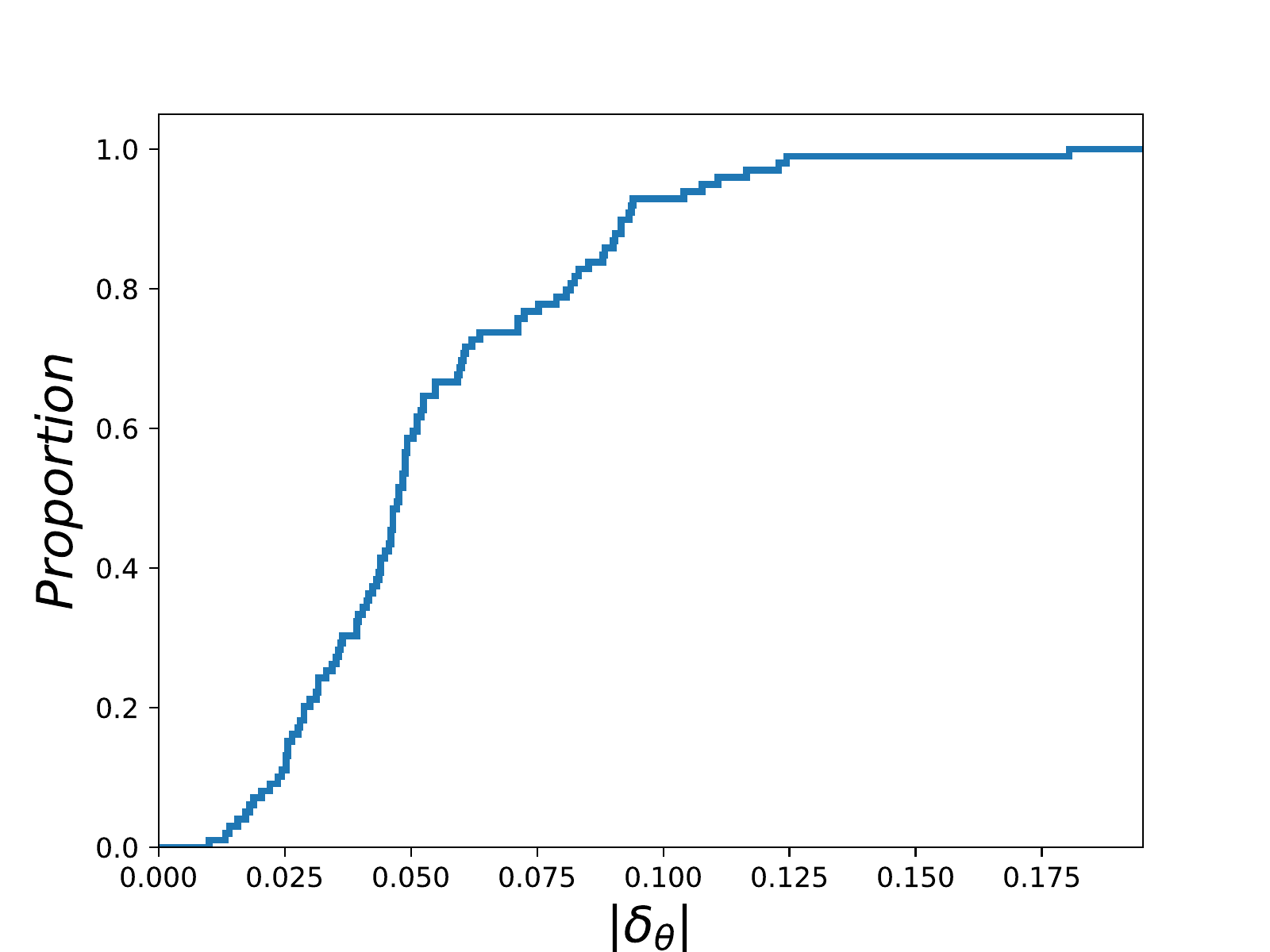}
\end{center}
\caption{\textbf{Cumulative probability distribution of $|{\delta _\theta }|$ for iSWAP-like gates.} $|\delta_{\theta}| $ has an average around $0.054$ and the speedup of SFA is less than an order.
\label{fSimgatesdistribution}}
\end{figure}

\begin{figure}[!htbp]
\begin{center}
\includegraphics[width=1.1\linewidth]{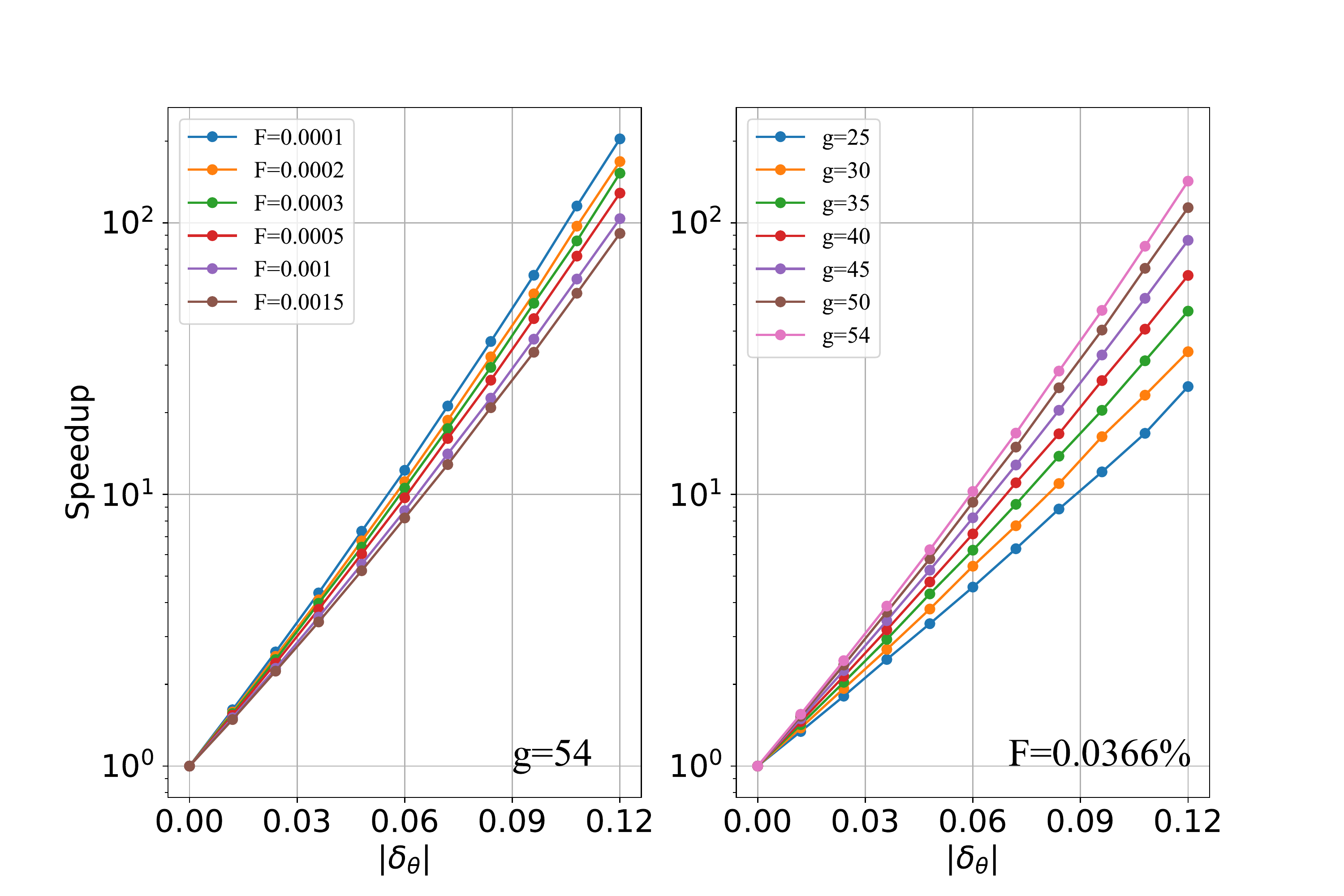}
\end{center}
\caption{\textbf{Classical speedup given by the imbalance gates.} We assume all iSWAP-like gates deviate from $\pi/2$ by the same $\delta_{\theta}$ and calculate the speedup with given $g$ and $F$. \textit{Left:} speedup with varied fidelity $F$ and fixed $g=54$. \textit{Right:} speedup with varied g and fixed $F=0.0366 \%$.
\label{imbalancedgates}}
\end{figure}

We did not consider the DCD formation in our estimation. The DCD formation appears three times in the 60-qubit 22-cycle and 24-cycle circuits. In addtion, the balanced iSWAP-like$(\theta ,\phi )$ gate in our experiment has  $\theta  \approx \pi /2$ and $\phi  \approx \pi /18$. Deviation of $\theta$ and $\phi$ from the expected values results in inbalanced gates, which can accerlerate the simulation with SFA. We calculate classical simulation speedup for imbalanced gates using the method introduced in Ref.~\cite{wu2021strong,arute2019quantum}.

Assuming the inbalanced gates in our experiments is in the form of iSWAP-like$(\pi/2 \pm {\delta _\theta },\phi  \approx \pi /18)$. We show the distribution of $|\delta _\theta|$ in Fig.~\ref{fSimgatesdistribution}. In the experiments, $|\delta _\theta|$ has an average 0.054 radians. For 60-qubit 24-cycle circuits, we have XEB fidelity $F = 0.0366\%$ and the number of decomposed
iSWAP-like gates $g = 54$ in SFA. The speedup resulted from inbalanced gates is shown in Fig.~\ref{imbalancedgates}. We can conjecture that the classical speedup in our experiment is less than an order.

\bibliography{references}